\documentclass[11pt]{article}
\pdfoutput=1
\usepackage{dcolumn}
\usepackage{bm}

\usepackage{slashed}
\usepackage[caption=false]{subfig}
\usepackage{jheppub}

\newcommand{\beq}{\begin{equation}}
\newcommand{\eeq}{\end{equation}}
\newcommand{\be}{\begin{equation}}
\newcommand{\ee}{\end{equation}}
\newcommand{\bea}{\begin{eqnarray}}
\newcommand{\eea}{\end{eqnarray}}

\title{Black Holes, Dark Matter Spikes, and Constraints on Simplified Models with $t$-Channel Mediators}

\author[]{Pearl Sandick$^{a}$,}

\author[]{Kuver Sinha$^{a,b}$,}

\author[]{and Takahiro Yamamoto$^{a}$}

\emailAdd{sandick@physics.utah.edu}
\emailAdd{kuver.sinha@ou.edu}
\emailAdd{t.yamamoto.1777@gmail.com}

\affiliation{$^a$Department of Physics and Astronomy, University of Utah, Salt Lake City, UT 84112, USA}
\affiliation{$^b$Department of Physics and Astronomy, University of Oklahoma, Norman, OK 73019, USA}

\abstract{

A possible density spike of dark matter (DM) in the subparsec region near the supermassive black hole at the Galactic Center can provide potentially observable gamma-ray signals emanating from DM annihilations. Taking Fermi-$LAT$ data for the gamma-ray flux from the point source  3FGL J1745.6-2859c (Sgr A$^*$), we calculate the resulting constraints on generic models of DM, allowing for the possibility of a non-negligible velocity-dependent component of the annihilation cross section.  We consider a range of values for relevant astrophysical parameters that describe the spike profile and find that the gamma-ray flux is strongly dependent on these choices; in particular, the modeling of spike depletion effects due to gravitational interactions with baryons, which affect the spike radius and the steepness of the profile. We consider both an idealized case where no attenuation of the spike occurs, as well as a case where the spike is depleted over time, and in each case we consider several choices for the steepness of the profile. We find that for the most conservative selection of parameters, corresponding to a depleted spike with an NFW cusp profile, the gamma-ray flux for a 100 GeV thermal relic is lower than current observational constraints by several orders of magnitude. For parameter choices corresponding to spikes that have not been attenuated, bounds on the mass of thermal DM can be obtained, and depend on the assumed steepness of the profile. We also specialize to a class of simplified models of fermionic DM that annihilate dominantly through the $t-$channel exchange of two scalar mediators with arbitrary mixing angle $\alpha$, and calculate the constraints on these models coming from the DM spike, for regions of parameter space that are complementary to collider searches. These simplified models demonstrate the sensitivity of conclusions about particle physics models to astrophysical parameters.  Finally, we discuss the possibility of constraining the astrophysical parameters describing the DM spike if the properties of the DM are known, taking as an example a proposed DM explanation for the observed excess of GeV photons from the GC region.

}

\begin{document}
\maketitle


\section{\label{sec:level1}Introduction}

The particle nature of dark matter (DM) is an area of intense investigation which has the potential to shed light on fundamental questions about the Standard Model (SM), especially the hierarchy problem. For DM candidates with weak-scale couplings and mass, a calculation of the relic density automatically yields a value that is close to the measured dark matter abundance. This striking fact, a success of the so-called Weakly Interacting Massive Particle (WIMP) paradigm, reinforces the possibility that DM is deeply connected to new physics at the weak scale. The indirect detection of the products of DM annihilation or decay are one potentially fruitful way to investigate the properties of DM. Indeed, if DM annihilations occurred in the early Universe, it is possible that we could observe the products of annihilations occurring today.

Indirect detection of WIMPs in the Milky Way halo has been a major endeavor over many years. The gamma-ray flux $\Phi$ coming from WIMP annihilation is proportional to the line-of-sight integral of the square of the DM density,
\be
\Phi \, \sim \, \int{\rho^2(r) dr} \,\,.
\ee
Since the Galactic Center is expected to have a very high density of DM, it has been a much-studied source for indirect detection of DM.

The formation of black holes at the centers of DM halos, and in particular the supermassive black hole at the center of our Galaxy \cite{Genzel:2003cn,Schodel:2003gy}, can significantly modify the DM profile and affect the observed gamma-ray flux from that region. Gondolo and Silk showed \cite{Gondolo:1999ef} that if the black hole grows adiabatically at the center of a cusp with a power-law profile, 
\be
\rho(r) \sim r^{-\gamma_c} \,\,\,\,\,\,{\rm (cusp \,\, profile)},
\ee
a DM spike can form close to the black hole, with a density profile given by
\be
\rho(r) \sim r^{-\gamma_{sp}} \,\,\,\,\,\,{\rm (spike \,\, profile)},
\label{spikeinitial}
\ee
with $\gamma_{sp}>\gamma_c$.  
Such a spike causes an increase in $\Phi$ due to the enhanced density $\rho$ in Eq.~\ref{spikeinitial} at small radii. In fact, as $r\rightarrow 0$ the DM density profile diverges, but the divergence is cut off by the black hole horizon and smoothed near it due to the effects of DM annihilation.

The account above is an idealized case, since the DM spike can be destroyed or smoothed by various effects \cite{Ullio:2001fb,Merritt:2003qk, Bertone:2005hw,Gnedin:2003rj, Ahn:2007ty}. In galactic nuclei, stars have much larger kinetic energy than DM particles, and interactions between them cause DM to be heated up. The gravitational interaction between stars near the black hole and the DM spike can thus cause damping, which affects the spike parameters, including the power-law behavior and the spike radius. The astrophysical parameters that describe the DM spike are a topic of ongoing debate, with a fairly broad range of plausible possibilities.

The purpose of this paper is to investigate contributions of annihilations in the DM spike to the gamma-ray flux $\Phi$.  Specifically, we investigate different spike profiles (i.e.~spike formation histories) to determine whether the expectation of an enhanced signal due to the presence of a spike is robust.  We consider cases where the DM profile is an {\it idealized} spike, which has not changed significantly since its formation, as well as cases where gravitational interactions with baryons have caused the spike to be {\it depleted} over time. 

Furthermore, as has been discussed in~\cite{Amin:2007ir,Shelton:2015aqa}, the DM velocity dispersion can be significantly altered near the GC, where the gravitational influence of the black hole is substantial.  In this case, even DM models in which the annihilation cross section today is velocity-suppressed may lead to non-negligible gamma-ray signals from the GC, where the velocities can be large.  Here, we investigate a range of DM models with both velocity-independent as well as velocity-dependent contributions to the annihilation cross section.  Finally, we present a concrete example of a model in which the conclusions from gamma-ray data depend strongly on the details of the DM spike: a simplified model of fermionic DM coupled to Standard Model fermions via charged scalars \cite{Sandick:2016zut}, \cite{Kumar:2016cum}.  

Indirect detection of DM from a spike near the central black hole of our galaxy has been studied  by several authors in different contexts in  particle physics. Recently, \cite{Fields:2014pia,Shelton:2015aqa,Shapiro:2016ypb} have studied these issues in the context of the Galactic Center excess and for DM models with $p$-wave annihilation for an idealized spike. Indirect detection of DM with a velocity-dependent annihilation cross section has been studied by \cite{Amin:2007ir}, in models of non-thermal DM by \cite{Sandick:2011rp}, and in the context of dark stars by \cite{Sandick:2010qu,Sandick:2010yd,Schoonenberg:2016aml}. Meanwhile, spikes at the center of dwarf galaxies have recently been constrained by \cite{Wanders:2014xia, Gonzalez-Morales:2014eaa}. 

Here, we expand on the studies of DM annihilation near the Milky Way GC.  We find that the size of the spike, denoted by the {\it spike radius}, and the steepness of the profile both inside the spike, parametrized by $\gamma_{sp}$, and outside (in the cusp), parametrized by $\gamma_c$, have a strong effect on the resulting constraints on DM models. For convenience, we summarize our main results: 

$(i)$ The most conservative choice of parameters, corresponding to an attenuated spike radius, an NFW profile for the DM cusp $\gamma_c = 1.0$, and a flattened annihilation core yields a flux $\Phi$ that is several orders of magnitude below the current observational threshold for a 100 GeV DM thermal relic (see the top left panel of Fig.~\ref{coc1resultsm100}).

$(ii)$ For a less conservative choice of parameters, corresponding to an attenuated spike radius, but a steeper profile for the DM cusp, $\gamma_c = 1.1-1.5$, thermal relics of various masses may be constrained as shown in  Fig.~\ref{fluxcomparedobserva}. The constraints on the velocity independent and velocity dependent contributions to the DM annihilation cross section are plotted in the top right and bottom panels of Fig.~\ref{coc1resultsm100} and Fig.~\ref{coc1resultsm200}.

$(iii)$ Assuming that the spike has not undergone depletion improves the constraints considerably. In this idealized case, one can constrain thermal relics of different masses as shown in Fig.~\ref{fluxcomparedobservb}, which displays various choices of $\gamma_c$, and assumes the steepest inner spike profile $\gamma_{sp}$ that one might reasonably expect. This steepest choice corresponds to a spike formed by collisionless DM assuming adiabatic growth of the central black hole.

$(iv)$ We also consider whether $\gamma_{sp}$ might be smaller than the steepest reasonable expectation, allowing it as a free parameter.
The results are more conservative than Case $(iii)$, and are displayed in Fig.~\ref{fluxcomparedobservc} for one choice of $\gamma_{sp}$ and a range of values of $\gamma_c$.

Our work suggests that a more careful study of the astrophysics of DM spikes near black holes, specifically in the neighborhood of the supermassive black hole (SMBH) at the center of our Galaxy, is warranted. The wide range of plausible spike parameters results in significant variation in the space of DM constraints. To illustrate this, we take the above cases and apply them to a simplified model, with results that are depicted in Fig.~\ref{simpc0c1adepalp} - Fig.~\ref{alpdepc0c1}.

The paper is structured as follows.  In Section \ref{spikephysics}, we discuss the parameters that describe the DM spike near the black hole. In Section \ref{results}, we discuss our main results in general DM models. In Section \ref{simpresults}, we describe our results in the context of a simplified DM model, in which dark matter annihilates via $t$-channel exchange of charged mediators. In Section \ref{spikeconstras}, we briefly discuss what can be learned about the DM spike under the assumption of a particular DM model, in this case one designed to explain the excess of GeV photons from the GC~\cite{Daylan:2014rsa}, \cite{Karwin:2016tsw}, \cite{TheFermi-LAT:2015kwa}. We end with our Conclusions.

\section{Dark Matter Spike Near the Supermassive Black Hole} \label{spikephysics}

In this section, we discuss the profile of a DM spike near the SMBH in the inner subparsec region of our Galaxy. This type of DM spike has been studied by many groups, beginning with the work of Gondolo and Silk \cite{Gondolo:1999ef}. 
In the following, we remain agnostic about the nature of DM, and parametrize its annihilation cross section as \cite{Srednicki:1988ce}
\begin{equation}\label{eq:partialwave}
\left \langle \sigma v \right \rangle \sim  c_0 + c_1 \left ( \frac{v}{c} \right )^2 \,\,,
\end{equation}
where $c_0$ is the velocity-independent $s$-wave contribution, and $c_1$ is the $v^2$-suppressed contribution. We note that the velocity-suppressed terms arise from both $s$-wave and $p$-wave matrix elements. 

\begin{figure}[t]
\centering
\includegraphics[width=.8\textwidth]{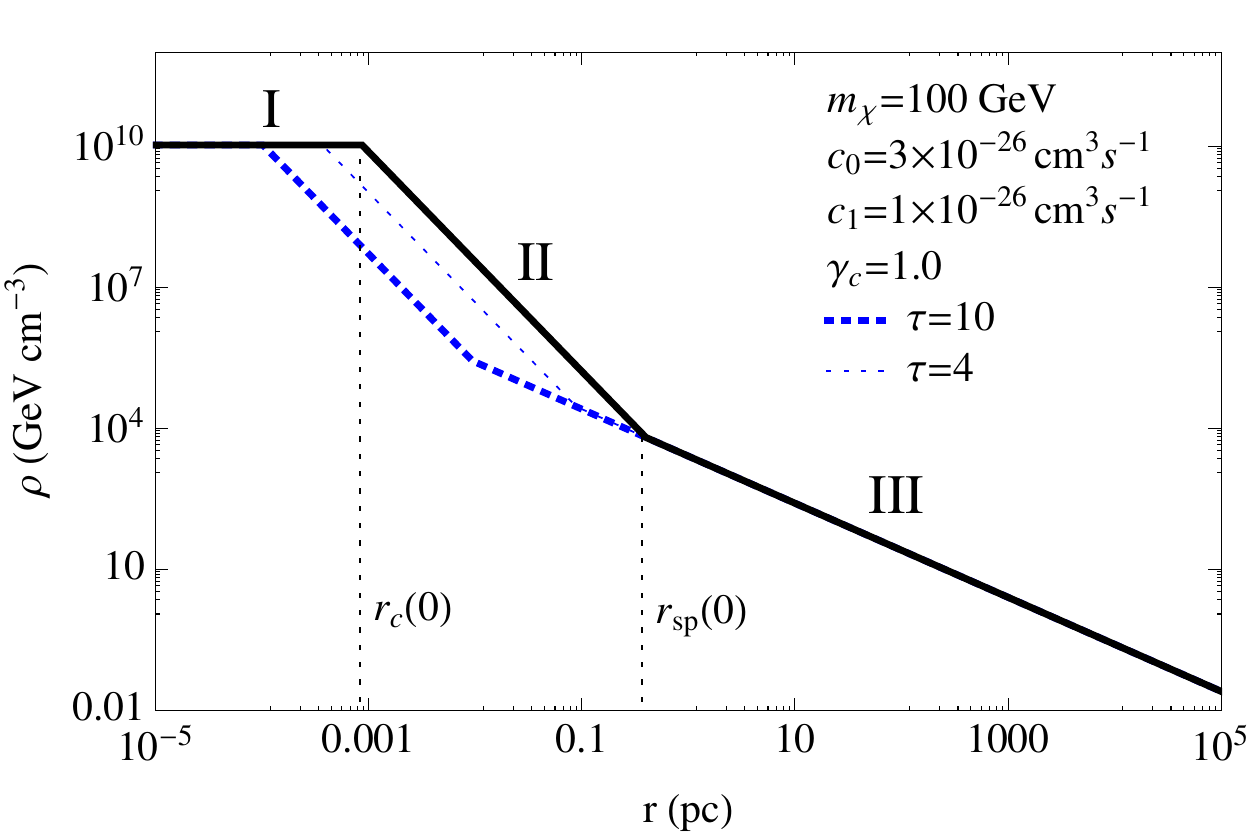}
\caption{The DM profile is displayed for a typical choice of parameters $\gamma_c = 1.0$, $\gamma_{sp} = 7/3$. The solid black profile corresponds to an idealized spike. The thick and thin blue dashed profiles correspond to depleted spikes with $\tau = 10$ and $\tau = 4$, respectively, where $\tau$ denotes the time since the spike formed in units of the heating time (details in text). }

  \label{profile}
\end{figure}

We consider a SMBH at the center of our Galaxy~\cite{Genzel:2003cn} with mass, $M_{bh}$, and Schwarzchild radius, $r_{Sch.}$, 
\bea
M_{bh} &=& 4 \times 10^6 M_{\odot} \nonumber \\
r_{Sch.} &\sim & 4 \times 10^{-7} \, {\rm pc} \,\,.
\eea
If the growth of the SMBH was adiabatic, and assuming collisionless dark matter particles, one finds that an original DM cusp with density profile $\rho(r) \sim r^{-\gamma_c}$ becomes contracted into a spike with profile $\rho(r) \sim r^{-\gamma_{sp}}$ at small radii~\cite{Gondolo:1999ef,Ullio:2001fb}.  In fact, at the smallest radii, just outside $r_{Sch.}$, the DM density likely attains a maximum or plateau value.  There are thus three distinct regions of the DM density profile, which will be discussed in detail in Section~\ref{sec:spikeprofile}.  Specifically, the profile is given by the analytic form
\begin{equation}
\label{density}
\rho(r)=\left\{ \begin{array}{lll}
\rho(r_{core}) & \qquad 10r_{Sch.}<r\le r_{core} & \qquad  (\textrm{Region III}),\\
 \rho_0 \left({r/r_{sp}}\right)^{-\gamma_{sp}}  & \qquad r_{core} < r
    \le  r_{sp} & \qquad  (\textrm{Region II}),\\
 \rho_0 \left({r/r_{sp}}\right)^{-\gamma_c} & \qquad r_{sp}
    < r & \qquad  (\textrm{Region I}).
\end{array} \right. 
\end{equation}
Here, $r_{sp}$ and $r_{core}$ denote the spike and core radii, respectively. The profile depends on the steepness parameters $\gamma_{sp}$ and $\gamma_c$.  Three example profiles are shown in Fig.~\ref{profile}.  

The formation of a DM spike is contingent on several conditions, detailed for example in \cite{Gondolo:1999ef} and \cite{Ullio:2001fb}.  After its formation, the DM spike may be dampened due to gravitational interactions with 
stars near the GC, or disrupted due to halo mergers, either of which can substantially reduce the steepness of the spike~\cite{Ullio:2001fb}. Here we consider the case of an {\it idealized} (undepleted) spike, as well as a spike that has been {\it depleted} due to gravitational interactions with stars.  For the latter, we follow the parametrization of \cite{Ahn:2007ty}.

We first give details about the spike and core radii, $r_{sp}$ and $r_{core}$, then we describe the physics of the profile for each of the three regions.

\subsection{Spike Radius ($r_{sp}$) and Depletion Effects}

In the idealized case, the spike radius does not evolve in time and is given by 
\be \label{rbinit}
r_{sp}(t) \, = \, r_{sp}(0) \, \sim \, 0.2 r_{h} \,\,\,\,\,\,\,\textbf{(Idealized \,\, Case)}.
\ee
Here, $r_{h}$ denotes the radius of gravitational influence of the black hole, 
\be
r_h \,\equiv\, {G M_{bh} \over \sigma^2} \,\,, 
\label{rhvalue}
\ee
where $G$ is Newton's constant and $\sigma$ denotes the one-dimensional velocity dispersion. $M_{bh}$ is related to $\sigma$ through the empirical relation~\cite{Ferrarese:2004qr}
\be
\frac{M_{bh}}{10^8 \, M_\odot} \,=\, (1.66 \pm 0.24) \, \left( { \sigma 
\over 200 \, \textrm{km s}^{-1} } \right)^{4.86 \pm 0.43} \,.
\label{rhvalue2}
\ee
%
For the central values in Eq.~\ref{rhvalue2}, we obtain 
\bea
\sigma \, \sim \, 93 \,\, {\rm km}/{\rm s} \nonumber \\
r_h \, \sim \, 1.99 \,\,\,{\rm pc} \,\,
\eea
for the DM velocity dispersion and the radius of influence of the black hole. In the idealized case, this leads to a spike radius of
\be
r_{sp} \, = \, 0.40 \,\,\,{\rm pc} \,\,\,\,\,\,\,\textbf{(Idealized \,\, Case)} \,,
\label{idealized}
\ee
denoted as $r_{sp}(0)$ in Fig.~\ref{profile}.

Gravitational interactions between DM and baryons will lead to changes from the idealized case of Eq.~\ref{rbinit}. Stars in the galactic nucleus have much larger kinetic energies than the DM particles, and the interactions between the two tend to heat up the DM. This leads to a dampening of the spike~\cite{Merritt:2003qk}. The decay of the spike can be described roughly as 
\be \label{eq:rhodamp}
\rho(r,t) \,\approx\, \rho(r,0) \, e^{-\tau/2} \,,
\ee
where $\tau$ is the time since the spike formed in units of the heating time, $t_{\rm heat}$~\cite{Ahn:2007ty}. The heating time is $t_{\rm heat} \approx 10^9$ years, and we will take $\tau = 10$ \cite{Bertone:2005xv}, though we note that these values are not well constrained.

The evolution of the spike radius can be described as~\cite{Ahn:2007ty}
\bea 
\label{depletion}
r_{sp}(t) &=& r_{sp}(0) \times \exp \left( \frac{-\tau}{2(\gamma_{sp} - \gamma_c)} \right) \,\,\,\,\,\,\, \textbf{(Depleted \,\, Case),} 
\eea
where $r_{sp}(0)=0.2r_h$ is the initial value of the spike radius.  Typical values for the depleted spike radius are
\bea
r_{sp} & = & 0.0094 \,\,{\rm pc} \,\,\,\,{\rm for } \,\,\,\, \gamma_c = 1.0   \nonumber \\
r_{sp} & = & 0.0015 \,\,{\rm pc} \,\,\,\,{\rm for } \,\,\,\, \gamma_c = 1.5\,\,\,\,\,\,\,\textbf{(Depleted \,\, Case)},
\eea
assuming a relation between $\gamma_{sp}$ and $\gamma_c$ as in Equation~\ref{gammaspeq}, described in Section~\ref{sec:spikeprofile}.
Note that the spike radius in the depleted case is much smaller than in the idealized case, and also that there is significant variation in the spike radius depending on $\gamma_c$. 
In Fig.~\ref{profile} we show an idealized spike (solid black), as well as two examples of depleted spikes, one with $\tau=10$ (think blue dashed) and one with $\tau=4$, i.e.~$t_{heat}=2.5\times 10^9$, (thin blue dashed).  Even for $\tau=4$, we see that the spike is significantly depleted.
Spikes of such varying size lead to a large range of predicted indirect signals of DM annihilation.
Throughout the rest of this study, we restrict our attention on the case of $\tau=10$ to demonstrate the effects of depletion.

\subsection{Dark Matter Density Profile}
\label{sec:spikeprofile}

In this section, we discuss the halo profile of the DM spike, starting from the outermost region and going to the innermost region.

\textbf{Region I - Outside the spike radius $r_{sp}$}: Typical DM halo profiles, such as NFW, are characterized by a double power law profile.  For the Milky Way, at radii less than $\mathcal{O}(10)$ kpc, the profile is a single power law, which we take to be relevant for $r>r_{sp}$, the radius inside which the spike is significant.  We may therefore parametrize the DM profile in Region I as
\be
\rho (r) \, = \, \rho_0 \left(\frac{r}{r_{sp}}\right)^{-\gamma_c}\,\,\,{\rm for}\,\, r_{sp} < r \,\,.
\ee
The normalization of the density profile $\rho_0$ is set by extrapolating inwards from the solar radius
\be
\rho_0 = \rho_{\odot} \left(\frac{r_{\odot}}{r_{sp}}\right)^{\gamma_c} \,\,,
\ee
where we take the density at the solar radius to be $\rho_{\odot} = 0.3$ GeV/cm$^{3}$.

N-body simulations that include only DM (and no baryonic matter) generally favor inner slopes of $\gamma_c \sim 1$, which is the canonical NFW value. However, baryonic interactions affect the profile in the inner 10 kpc of our galaxy, and can substantially steepen the power-law behavior \cite{Diemand:2008in, Navarro:2008kc, Gnedin:2004cx, Gustafsson:2006gr, Pato:2015dua}. 
Furthermore, observations are compatible with $\gamma_c$ as large as at least 1.5~\cite{Pato:2015dua}.
Here we consider a range of cusp exponents, allowing values of $\gamma_c\in [1.0,1.5]$.

\textbf{Region II - Inside the spike radius $r_{sp}$ but outside the core radius $r_{core}$}:  The spike profile itself is also parametrized as a simple power law,  
\be \label{densityii}
\rho (r) \, = \, \rho_0 \left(\frac{r}{r_{sp}}\right)^{-\gamma_{sp}}\,\,\,{\rm for}\,\, r_{core} < r \leq r_{sp} \,\,,
\ee
where the spike slope, $\gamma_{sp}$ may or may not be directly related to the cusp slope, $\gamma_c$. For collisionless DM forming a spike due to the adiabatic growth of the black hole, the spike slope obeys the relation 
\be
\gamma_{sp} \, = \, \frac{9-2 \gamma_c}{4 - \gamma_c} \,\,,
\label{gammaspeq}
\ee
which yields a value $\gamma_{sp} \approx 2.3 \, - \, 2.4$ for $1.0 \leq \gamma_c \leq 1.5$.  This relation holds for a central black hole that grows adiabatically from a small seed. 

However, the spike slope may be significantly different than the adiabatic expectation under different assumptions. If the black hole appeared instantaneously, then one obtains $\gamma_{sp} = 4/3$ \cite{Ullio:2001fb}. Black hole mergers at the center of the progenitor halo can give $\gamma_{sp} = 1/2$, a value that is also obtained if the black hole grows away from the center of the DM distribution \cite{Ullio:2001fb}. As above, we focus on the effect of stellar heating, which could result in a final equilibrium value as low as $\gamma_{sp} \sim 1.5$~\cite{Merritt:2003qk, Gnedin:2003rj}.
Note that more recent work by the author of~\cite{Merritt:2003qk}, namely~\cite{Ahn:2007ty}, indicates that the effect of stellar heating will be a decrease in $r_{sp}$, rather than a direct decrease in $\gamma_{sp}$ with $r_{sp}$ unchanged.  In the remainder of the paper, we primarily follow~\cite{Ahn:2007ty}, though we also address the possibility of a reduced value of $\gamma_{sp}$ relative to the adiabatic expectation.  In the latter case, we choose $\gamma_{sp}=1.8$, following~\cite{Fields:2014pia}.


\textbf{Region III - Inside the core radius $r_{core}$}: At very small radii, the DM density can reach very high values. However, that implies large values of the annihilation cross section, which acts to reduce the density. We make the conservative assumption that an annihilation plateau is formed in this region, with
\be
\rho (r) \, = \, \rho(r_{core}) \,\,\,{\rm for}\,\, 10 \, r_{Sch.} < r \leq r_{core}\,\,,
\ee
where the relevant inner radius is related to the Schwarzschild radius of the black hole, and the outer radius is the core radius, defined by the relation
\be \label{coreraddef}
\frac{\rho (r_{core})}{m_\chi} \langle \sigma v \rangle \, \sim \, (\tau t_{\rm heat})^{-1}\,\,,
\ee
%
which depends on the thermally averaged annihilation cross section times velocity, $\langle\sigma v\rangle$, and the WIMP mass, $m_\chi$.

We note that in the general case of arbitrary velocity anisotropy, instead of circular particle orbits, a cusp with $\rho \propto r^{(-\beta -1/2)}$ may develop in the center, where $\beta$ is the anisotropy coefficient \cite{Vasiliev:2007vh}. Though the cusp is expected to be very weak, it may further enhance the flux of DM annihilation products from the very central region of the Galaxy~\cite{Shapiro:2016ypb}. Here we take the simple limit of circular orbits, in which case there is a flat plateau as depicted in Fig.~\ref{profile}.

Finally,  we assume a virialized halo such that 
\be \label{vdepr}
\left(\frac{v}{c}\right)^2 = \frac{r_{Sch.}}{2r}\,\,.
\ee
Since the ratio $v/c$ appears in the partial wave expansion of the annihilation cross section, given by Eq.~\ref{eq:partialwave}, the annihilation cross section is therefore position-dependent, and the velocity-suppressed contribution can become large near the SMBH.  In fact, from Eq.~\ref{coreraddef}, we see that $r_{core}$, and therefore also $\rho(r)$ for $r<r_{core}$, may be sensitive to the velocity-suppressed contribution to the annihilation cross section, and will  in general vary with the coefficient $c_1$ even for fixed $c_0$.


\section{Results: Generic DM Model} \label{results}

In this section, we discuss our calculation of the gamma-ray flux from a DM spike at the Galactic Center, then we discuss the sensitivity of constraints on the properties of DM to assumptions about the form of the spike.  We begin by discussing depleted spikes, i.e.~those for which the spike density is dampened as in Eq.~\ref{eq:rhodamp}, which manifests as a decrease in the spike radius according to Eq.~\ref{depletion}.  In these cases, we assume the standard adiabatic relation for $\gamma_{sp}(\gamma_c)$, Eq.~\ref{gammaspeq}. 
Example profiles are shown as blue dashed contours in Fig.~\ref{profile}, and our results for depleted spikes are presented in Figs.~\ref{coc1resultsm100}-\ref{fluxcomparedobserva}. Next, we turn our attention to idealized spikes, which have not suffered a decrease in $r_{sp}$.  In this case we also consider both the adiabatic expectation for $\gamma_{sp}$, an example of which is shown as the solid black contour in Fig.~\ref{profile}, while our general results are presented in Fig.~\ref{fluxcomparedobservb}, as well as the effect of a direct decrease in $\gamma_{sp}$ with no change in spike radius, the results for which are presented in Fig.~\ref{fluxcomparedobservc}.

%
%

%
%




The differential flux of gamma rays from a given angular direction $d\Omega$ produced by the annihilation of Majorana DM, $\chi$, is given by
\bea
\label{gammaflux}
\frac{d \Phi_\gamma}{d\Omega\,dE} &=& \frac{1}{2}\frac{r_\odot}{4\pi} \left(\frac{\rho_\odot}{m_{\chi}}\right)^2   \int_{\rm l.o.s.} \frac{ds}{r_\odot} \left(\frac{\rho(r(s,\theta))}{\rho_\odot}\right)^2 \sum _f\langle \sigma v\rangle_f \frac{dN_\gamma^f}{dE}.
\eea
Here, $r(s,\theta)=(r_\odot^2+s^2-2\,r_\odot\,s\cos\theta)^{1/2}$ is the radial Galactic coordinate, and $\theta$ is the aperture angle between the direction of the line of sight, $s$, and the axis connecting the Earth to the Galactic Center. $dN_\gamma^f/dE$ is the spectrum of photons coming from annihilation to a final state $f$, and is computed with Pythia \cite{Sjostrand:2006za}. We note that the usual separation between the calculation of the astrophysical $J-$factor and the annihilation cross section is no longer applicable here, since the annihilation cross section itself depends on position, from Eq.~\ref{vdepr}.  

If the DM spike is a bright and compact enough source of photons, it may have been identified as a point source in Fermi-$LAT$'s Third Point Source Catalog. We will thus be interested in computing the total flux from the DM spike and comparing the brightness to point sources in the same region. As a comparative value that determines observability, we consider the integrated flux from 1 to 100 GeV for the Fermi 3FGL source J1745.6-2859c (Sgr A$^*$), which we denote as $\Phi_{\rm Fermi} = 2.18 \times 10^{-8}$ photons/cm$^2$s. If the flux from the spike exceeds this value, we consider the model to be excluded.  In fact, the bulk of the contribution to the gamma-ray flux from Sgr A$^*$ is from standard astrophysics, rather than DM annihilation, as a spectral analysis would reveal (see, eg.~\cite{Abazajian:2014fta}).  Here we are interested only in order of magnitude estimates, and prefer to remain agnostic about the nature of the DM, so we take the only constraint to be the upper limit on the integrated flux.  Furthermore, rather than assuming a specific final state to which dark matter annihilates, we choose as a benchmark value for the integrated photon count $N=1$, with the flux scaling $\propto N$, and integrate over a fixed angular acceptance of $0.1^\circ$\,\footnote{Choosing a small angular acceptance rather than using the full PSF may underestimate the flux from the spike by a factor of a few, depending on the final state.  Here, we prefer agnosticism regarding the annihilations themselves, and are interested in broad trends in detectability. We will see that the exponents $\gamma_c$ and $\gamma_{sp}$ can cause variations in the flux by several orders of magnitude.}.

\subsection{Depleted Spike}

   \begin{figure}[ht]
  \centering
  {\includegraphics[width=.45\textwidth]{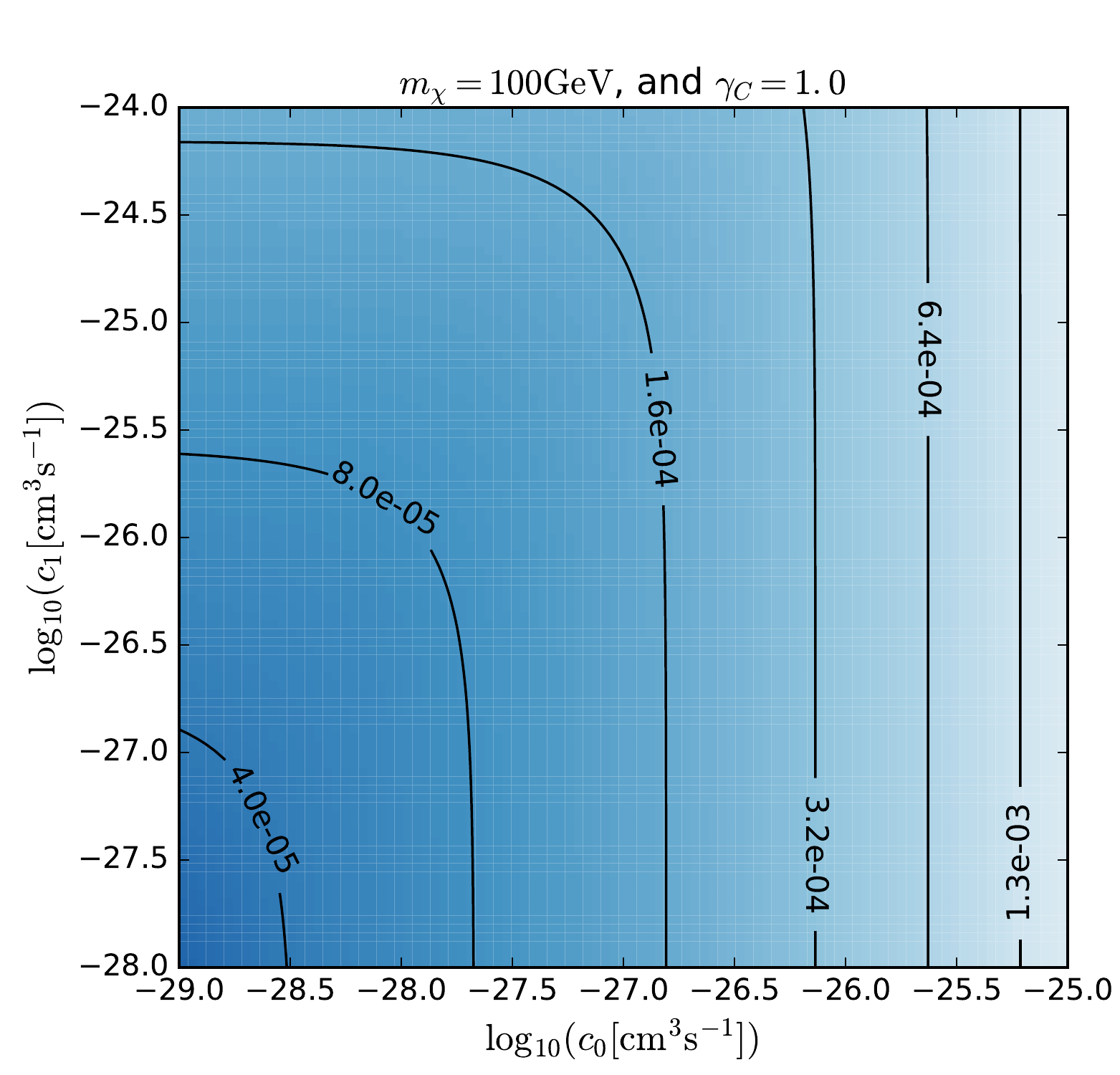}}\quad
  {\includegraphics[width=.45\textwidth]{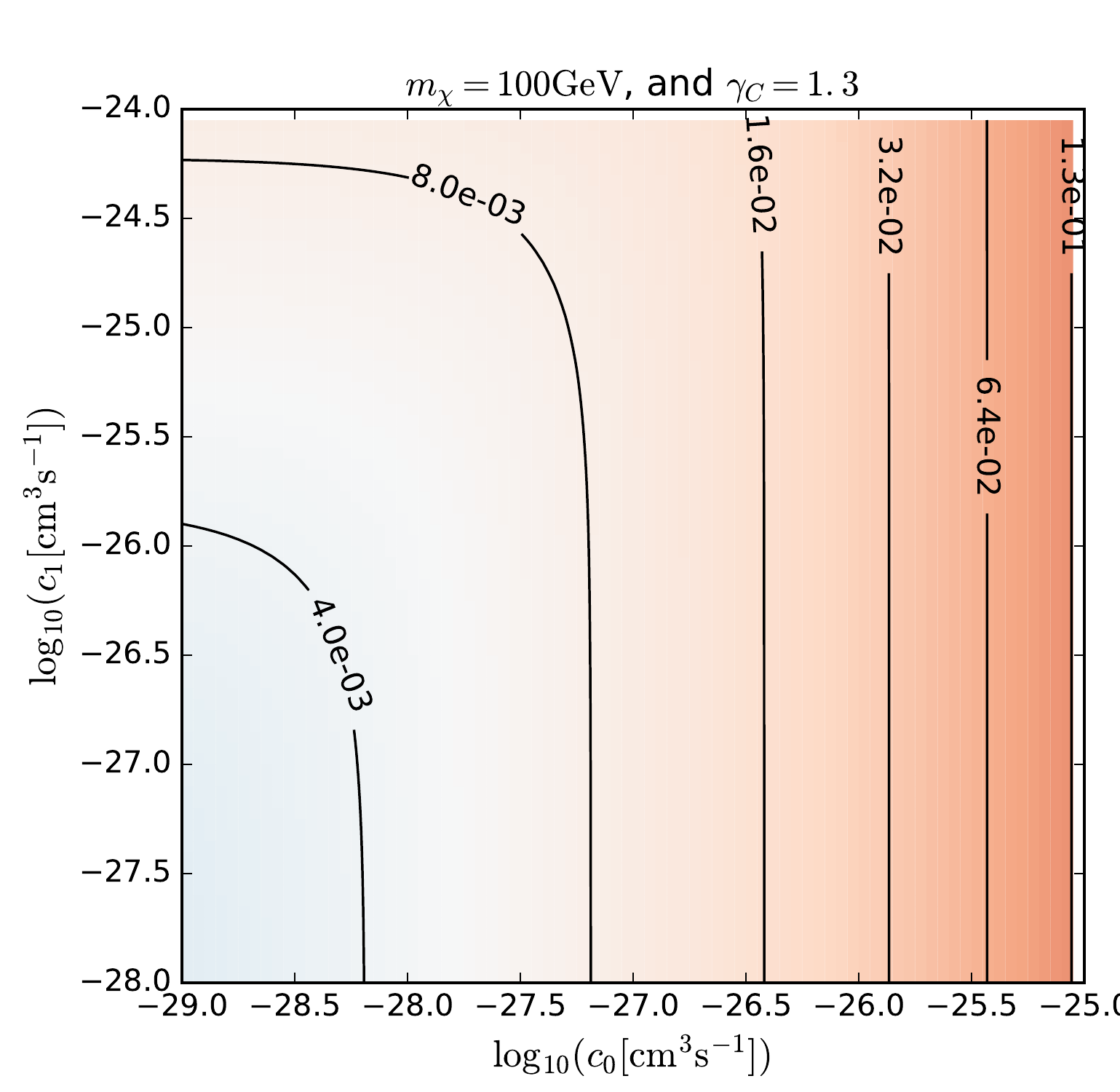}}\quad 
  {\includegraphics[width=.55\textwidth]{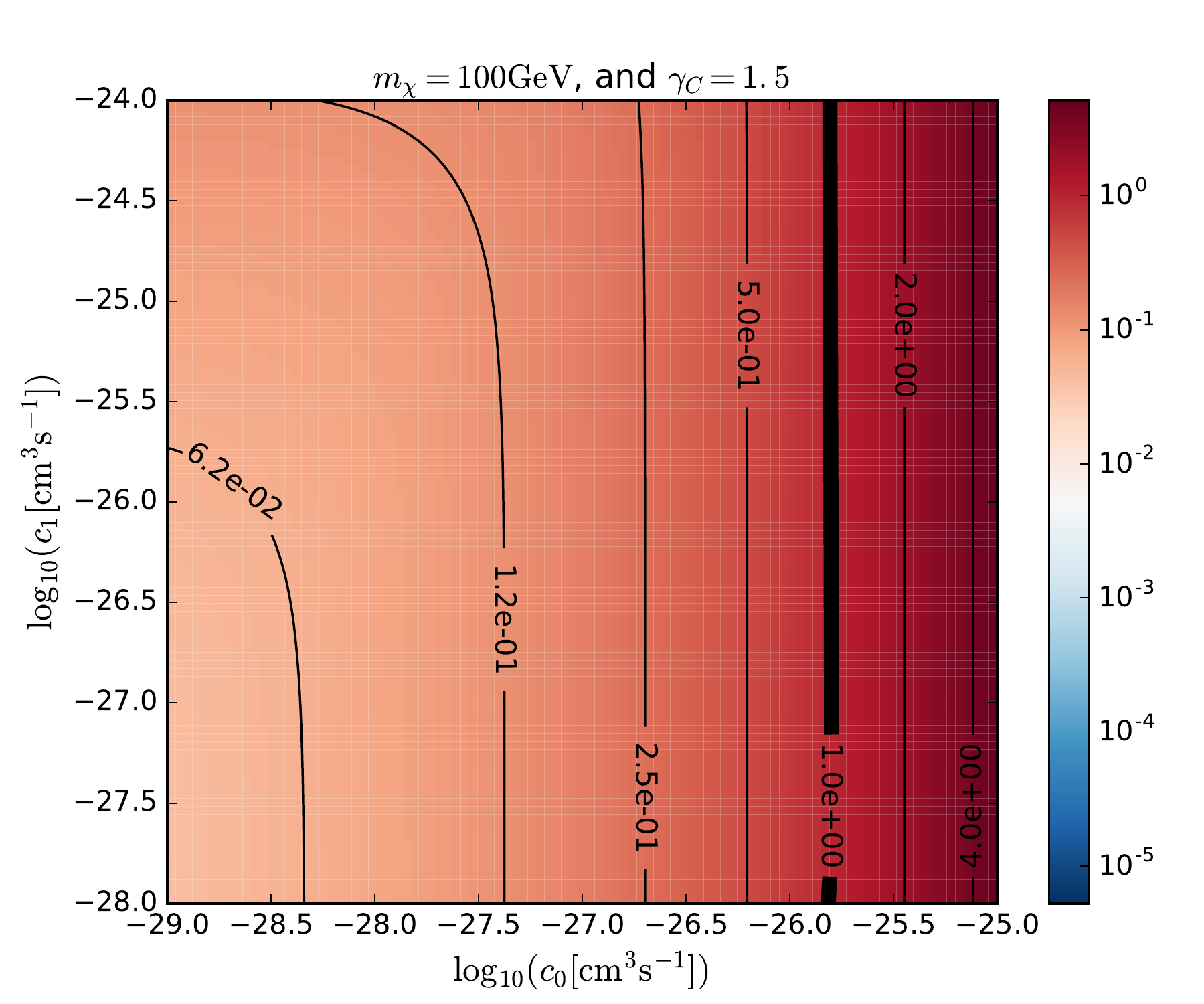}}
    \caption{\textbf{Depleted Spike,} $\mathbf{100}$ \textbf{ GeV DM: Contours of the integrated flux} $\mathbf{\Phi}$ in units of $\Phi_{\rm Fermi} = 2.18 \times 10^{-8}$ photons/cm$^2$s coming from the source 3FGL J1745.6-2859c (Sgr A$^*$), in the energy range 1-100 GeV, and assuming an integrated photon count $N=1$. The dark matter mass is 100 GeV, and the annihilation cross section is parametrized by Eq.~\ref{eq:partialwave}. The spike profile is given by Eq.~\ref{density}. The spike radius is given by the depleted case in Eq.~\ref{depletion}. The spike power law  outside the spike radius is given by $\gamma_c  = 1.0$ (upper left panel), $\gamma_c  = 1.3$ (upper right panel), and $\gamma_c  = 1.5$ (lower panel). The bold line in the lower panel shows the contour $\Phi = \Phi_{\rm Fermi}$. The spike power law inside the spike radius is given by Eq.~\ref{gammaspeq}, yielding values in the range $\gamma_{sp} \sim 2.3 \, - \, 2.4$ .}
    \label{coc1resultsm100}
\end{figure}

   \begin{figure}[ht]
  \centering
  {\includegraphics[width=.45\textwidth]{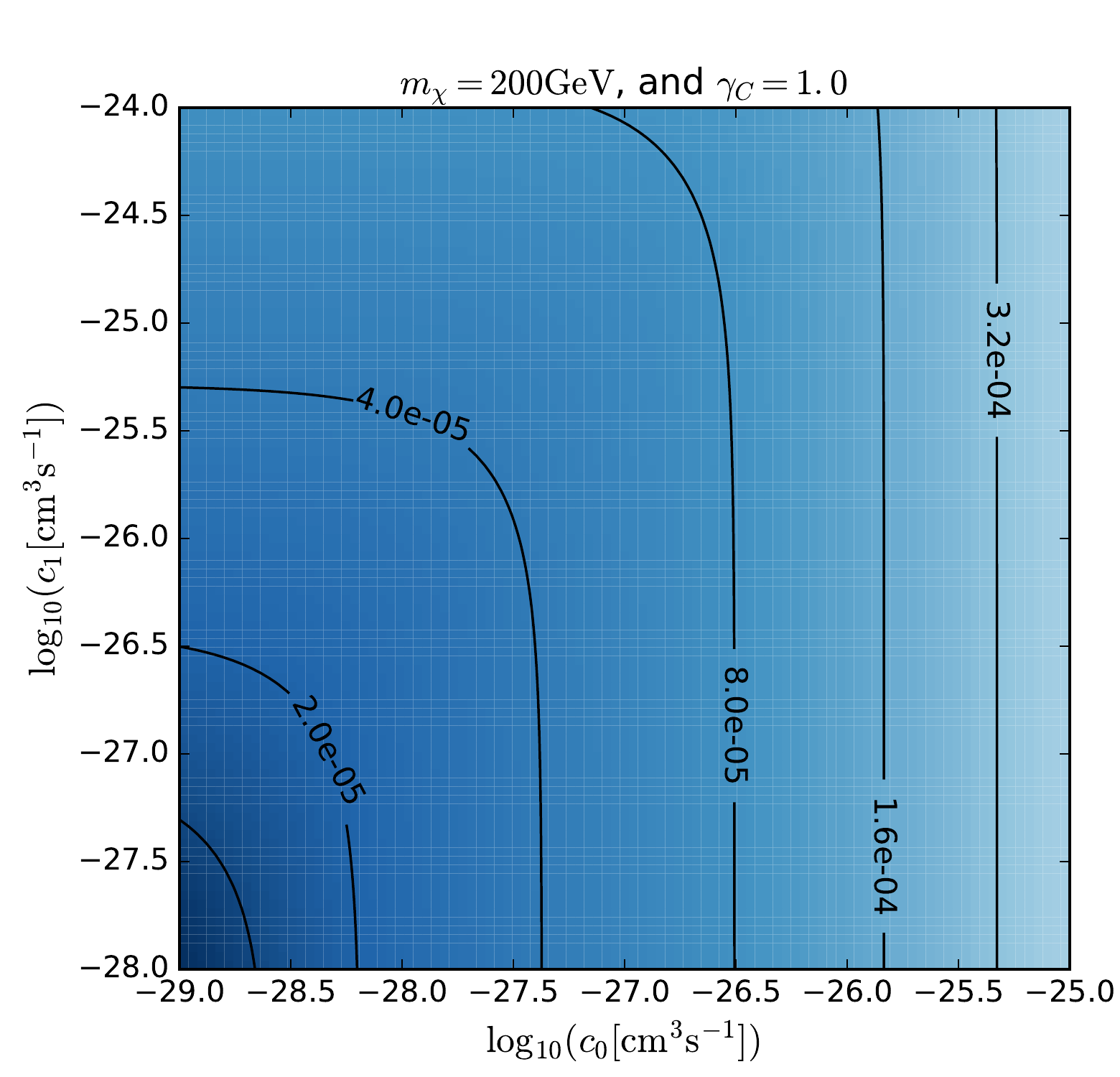}}\quad
  {\includegraphics[width=.45\textwidth]{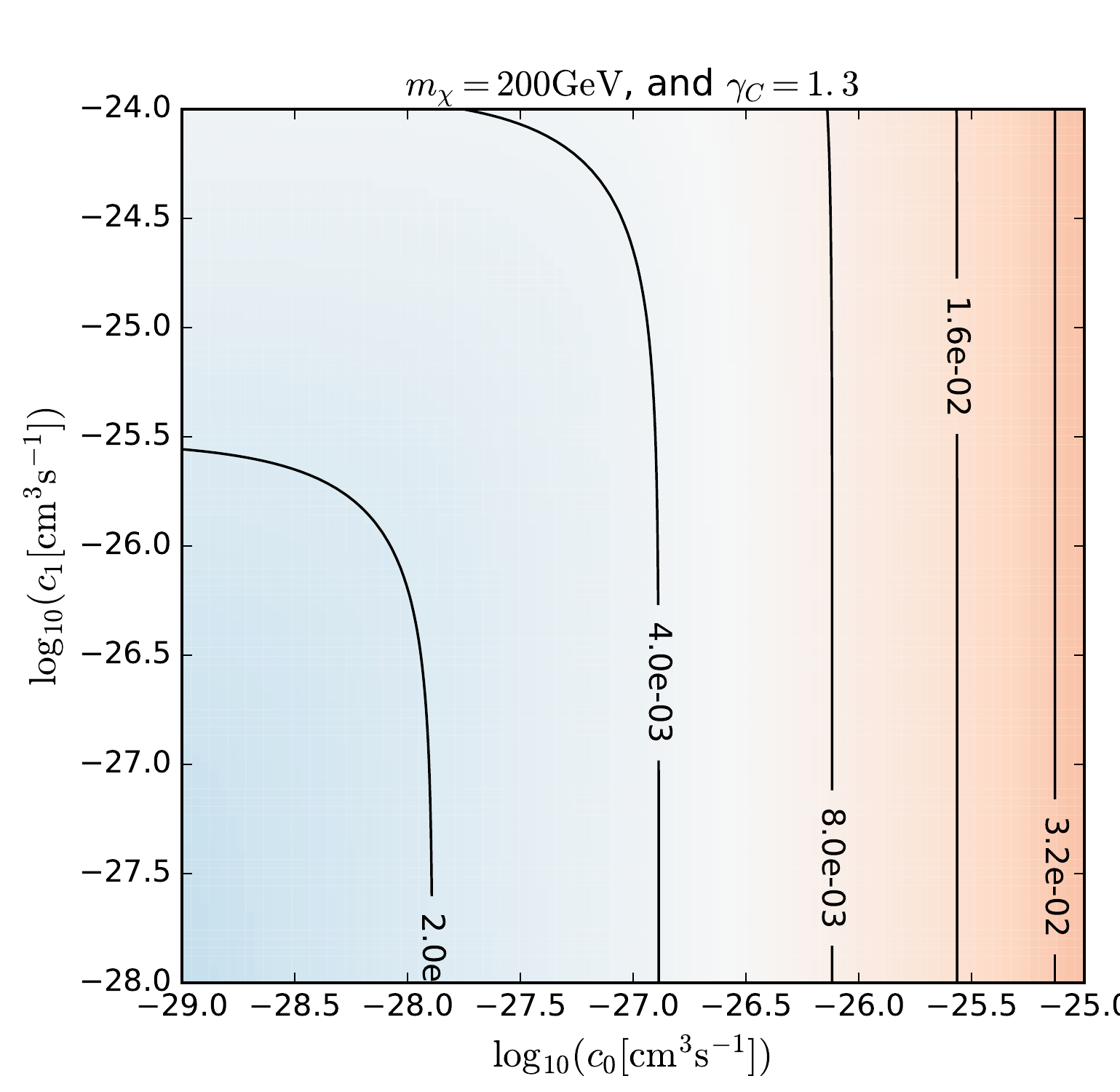}}\quad 
  {\includegraphics[width=.55\textwidth]{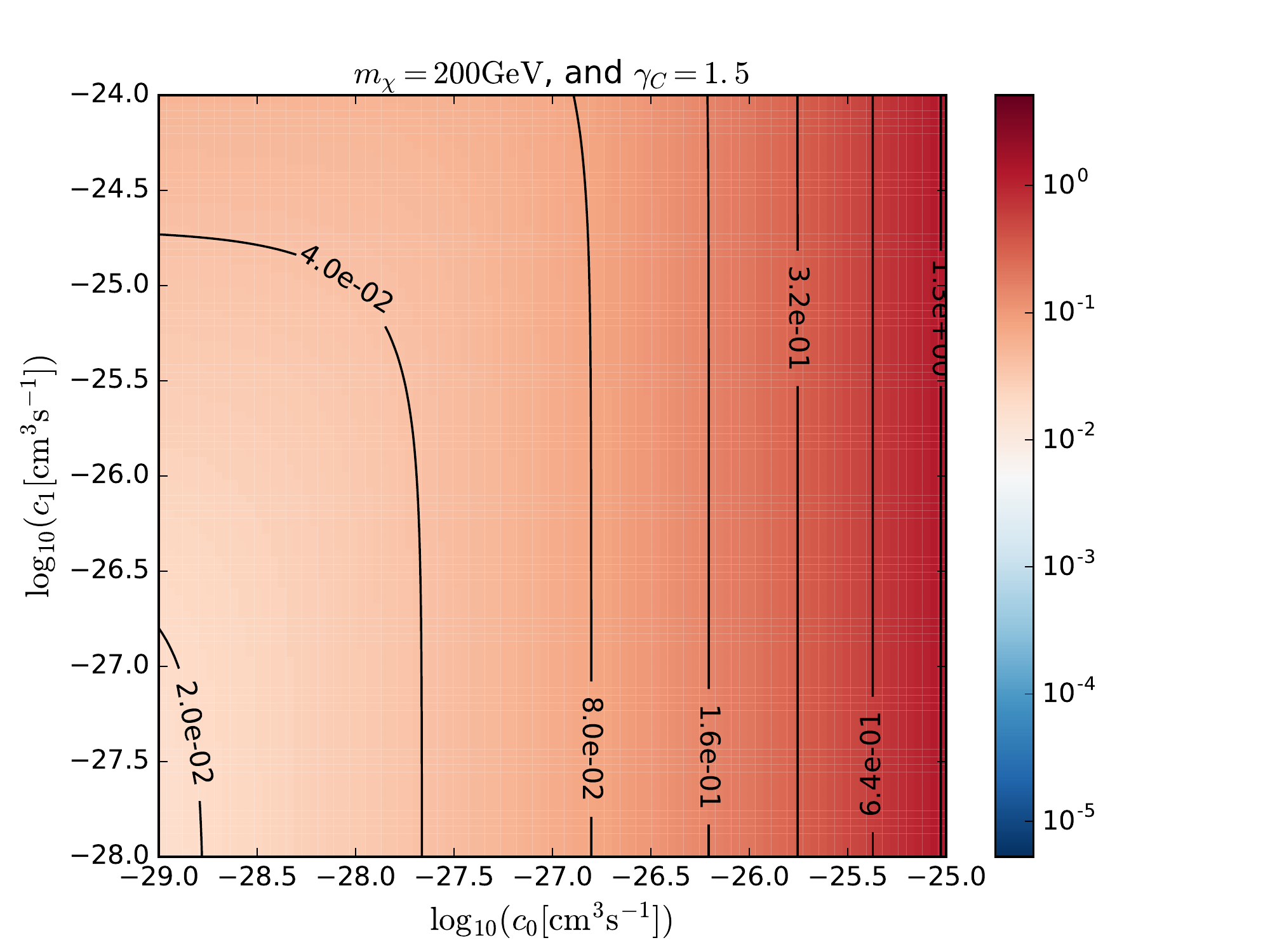}}
    \caption{\textbf{Depleted Spike,} $\mathbf{200}$ \textbf{ GeV DM: Contours of the integrated flux} $\mathbf{\Phi}$ in units of $ \Phi_{Fermi} = 2.18 \times 10^{-8}$ photons/cm$^2$s coming from the source 3FGL J1745.6-2859c (Sgr A$^*$), in the energy range 1-100 GeV, and assuming an integrated photon count $N=1$. The dark matter mass is 200 GeV, and the annihilation cross section is parametrized by Eq.~\ref{eq:partialwave}. The spike profile is given by Eq.~\ref{density}. The spike radius is given by the depleted case in Eq.~\ref{depletion}. The spike power law  outside the spike radius is given by $\gamma_c  = 1.0$ (upper left panel), $\gamma_c  = 1.3$ (upper right panel), and $\gamma_c  = 1.5$ (lower panel). The spike power law inside the spike radius is given by Eq.~\ref{gammaspeq}, yielding values in the range $\gamma_{sp} \sim 2.3 \, - \, 2.4$.}
    \label{coc1resultsm200}
\end{figure}

In Figures~\ref{coc1resultsm100} and~\ref{coc1resultsm200} we present our results in the $(c_0,c_1)$ plane for  $m_\chi=100$ GeV and  $m_\chi=200$ GeV, respectively, for the case of a depleted spike.  We explore a range of values of $\gamma_c = 1.0, \, 1.3,$ and 1.5. The contours show the total integrated flux $\Phi$ in units of  $\Phi_{\rm Fermi}$. In both figures, we see that for $\gamma_c \lesssim 1.3$, the total integrated flux $\Phi$ is below current observational sensitivity for the entirety of the parameter space shown, i.e.~$c_0 \lesssim 10^{-25}$ cm$^{3}$ s$^{-1}$ and $c_1 \lesssim 10^{-24}$ cm$^{3}$ s$^{-1}$. For $m_\chi=100$ GeV and $\gamma_c = 1.5$, the observed gamma-ray flux constrains models with annihilation cross section $c_0 \gtrsim 1.6 \times 10^{-26}$ cm$^{3}$ s$^{-1}$. Therefore a canonical thermal relic with cross section $3 \times 10^{-26}$ cm$^{3}$ s$^{-1}$ is excluded by this choice of spike parameters if its mass is $\lesssim100$ GeV.  In Fig.~\ref{coc1resultsm200} we see that the constraints weaken as the DM mass increases.

One can also see that if the velocity-dependent component $c_1$ provides the dominant contribution to the photon flux, it must be significantly larger (by a few orders of magnitude) than $c_0$ need be if it dominates. This is due to the factor $(v/c)^2 \sim (r_{Sch.}/2r)$, which is small away from the central black hole.  At the end of the day, the velocity-independent contribution to the annihilation cross section is likely still dominant.  Note, however, that it is conceivable that $c_0=0$, in which case a very large value of $c_1$ could lead to a signal from DM annihilation in the spike, when otherwise no indirect detection signal would be expected.

   \begin{figure}[ht]
  \centering
  {\includegraphics[width=\textwidth]{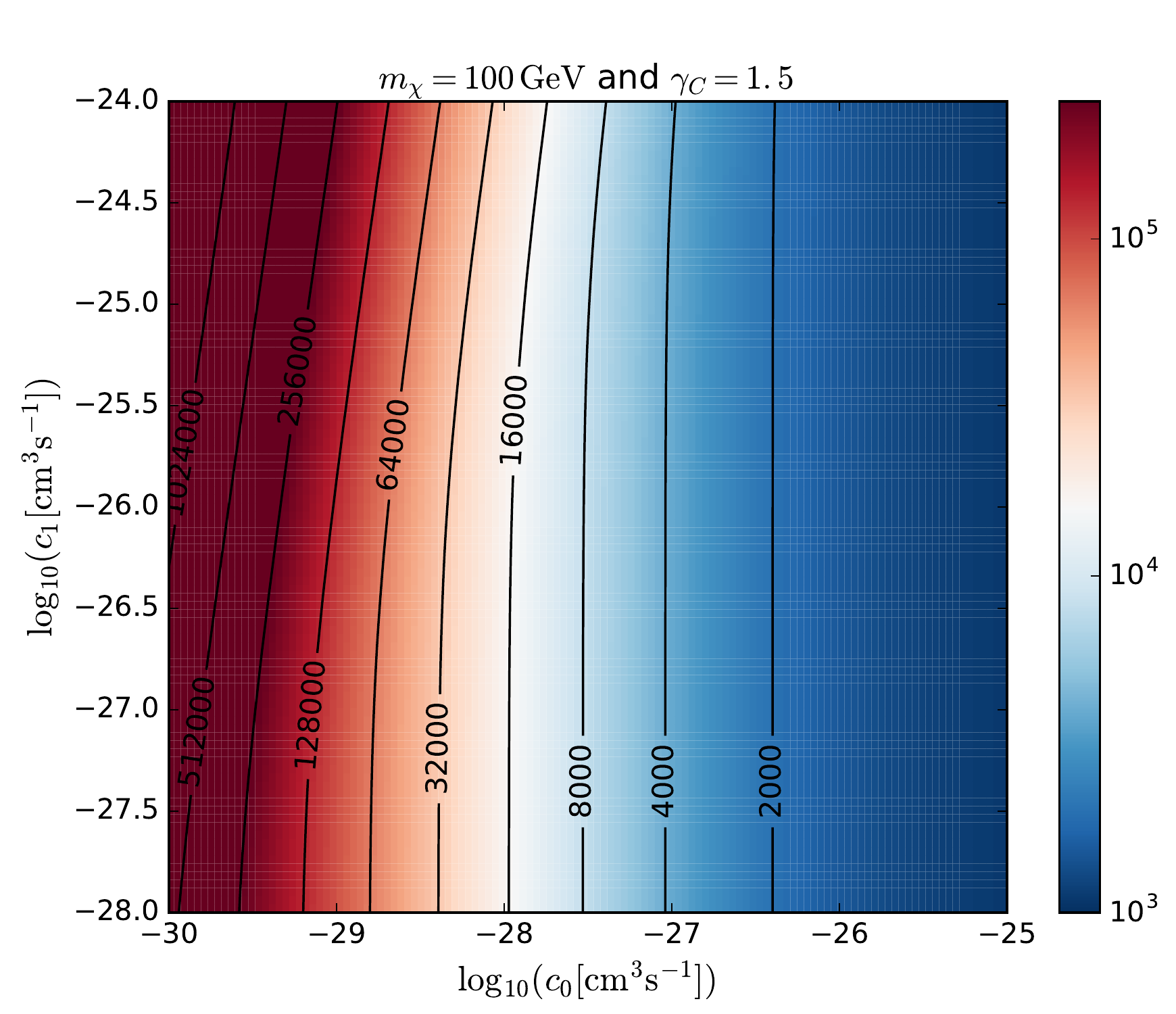}}
    \caption{\textbf{Depleted Spike,} \textbf{Contours of the ratio $\Phi/\Phi_{\rm{NFW}}$}, where $\Phi$ and $\Phi_{\rm{NFW}}$ are the total integrated flux in the presence and absence, respectively, of a DM spike near the supermassive black hole, coming from the source 3FGL J1745.6-2859c (Sgr A$^*$), in the energy range 1-100 GeV, and assuming an integrated photon count $N=1$. The dark matter mass is 100 GeV, and the annihilation cross section is parametrized by Eq.~\ref{eq:partialwave}. The spike profile is given by Eq.~\ref{density}. The spike radius is given by the depleted case in Eq.~\ref{depletion}. The spike power law  outside the spike radius is given by $\gamma_c  = 1.5$. The spike power law inside the spike radius is given by Eq.~\ref{gammaspeq}, with value $\gamma_{sp} \sim 2.4$.}
    \label{fluxcomparednfw}
\end{figure}

In Fig.~\ref{fluxcomparednfw}, we show contours of the ratio $\Phi/\Phi_{\rm{NFW}}$, where $\Phi$ and $\Phi_{\rm{NFW}}$ are the total integrated flux from annihilation of 100 GeV DM particles in the presence of a depleted spike with $\gamma_c=1.5$ and for a standard NFW profile ($\gamma_c=1.0$) with no spike. In the former case, the spike power law inside the spike radius is given by the standard adiabatic relation, Eq.~\ref{gammaspeq}. Unsurprisingly, we see that large enhancement factors are obtained relative to what one would expect for a standard NFW profile, even in the case of significant depletion as presented here.  This is due almost exclusively to the large cusp exponent, $\gamma_c$.  Indeed, for this value of $\gamma_c$, moderate values of $c_0$ and $c_1$ lead to an almost imperceptible spike (that is, $r_{sp} \approx r_{core}$).  However, even in this case, if $c_1\gg c_0$, then the impact of annihilations in the very central region of the spike becomes enhanced by the high velocities of the DM particles,.  In the figure, vertical contours indicate that the flux is independent of $c_1$, which one would expect for slow-moving/cold dark matter, but the contours deviate from vertical when the velocity-dependent contribution to the annihilation cross section becomes important.  

Fig.~\ref{fluxcomparednfw} demonstrates three features of our analysis:  1) Especially for large $\gamma_c$, extremely large variations from the DM annihilation flux one would expect from an NFW halo are possible.  2) These variations are expected even in the relative absence of a significant spike ($r_{sp}\approx r_{core}$).  And 3), as discussed in~\cite{Fields:2014pia}, the flux may be much larger than the velocity-independent expectation if $c_1$ is large enough.

   \begin{figure}[ht]
  \centering
  {\includegraphics[width=\textwidth]{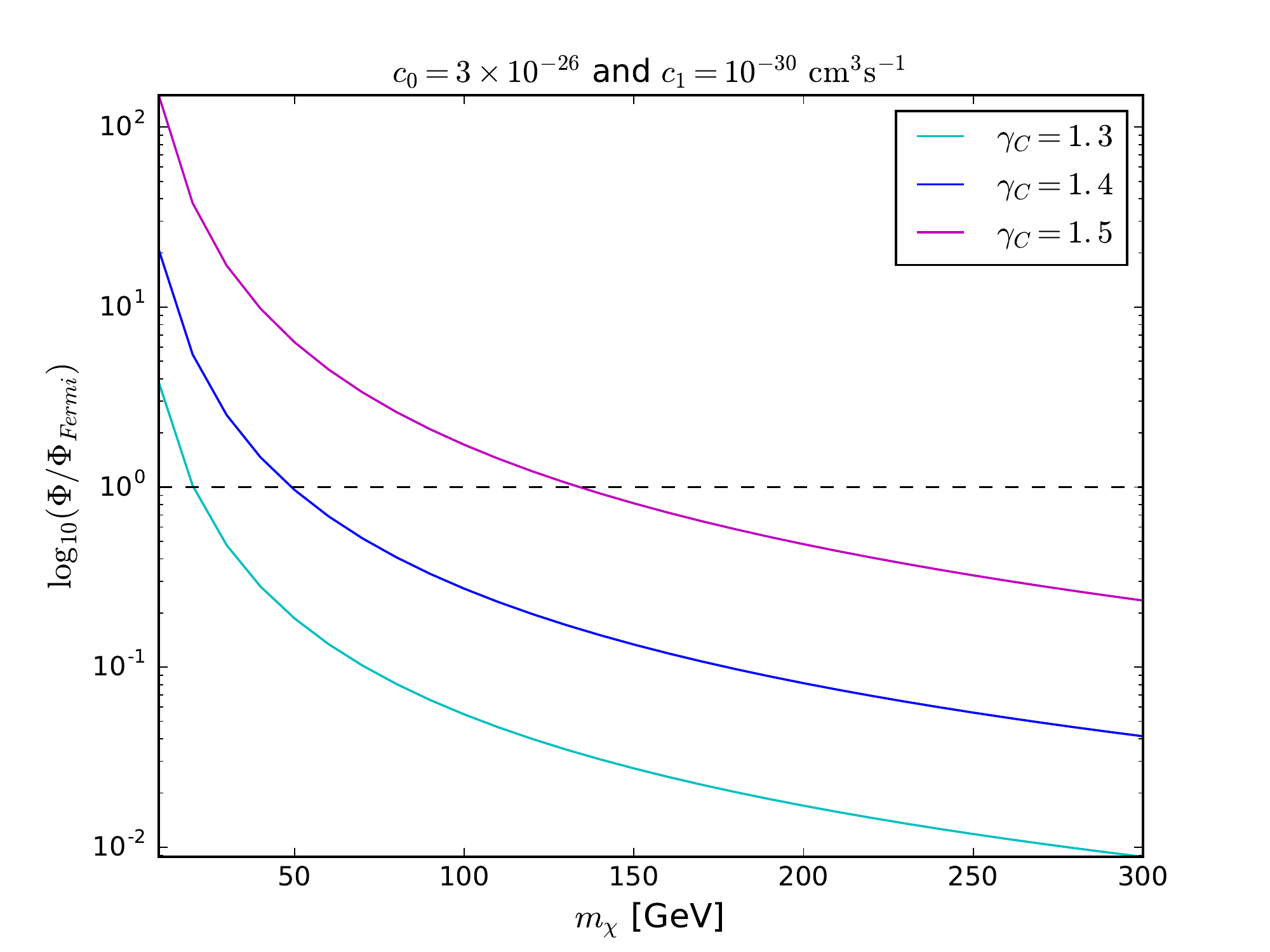}}
    \caption{\textbf{Depleted spike with} $\mathbf{\gamma_{sp} \, = \, \frac{9-2 \gamma_c}{4 - \gamma_c}} \, \sim \, 2.3 \, - \, 2.4$ \textbf{: Observational reach versus mass plot - } The total integrated flux $\Phi$ in units of  $ \Phi_{Fermi} = 2.18 \times 10^{-8}$ photons/cm$^2$s coming from the source 3FGL J1745.6-2859c (Sgr A$^*$), in the energy range 1-100 GeV, and assuming an integrated photon count $N=1$. The annihilation cross section is parametrized by Eq.~\ref{eq:partialwave}, with $c_0 = 3 \times 10^{-26}$ cm$^{3}$ s$^{-1}$ and $c_1 = 1 \times 10^{-30}$ cm$^{3}$ s$^{-1}$, corresponding to a canonical thermal relic. The spike profile is given by Eq.~\ref{density}. The spike radius is given by the depleted case in Eq.~\ref{depletion}. The spike power law inside the spike radius is given by Eq.~\ref{gammaspeq}, yielding values in the range $\gamma_{sp} \sim 2.3 \, - \, 2.4$. The dotted line shows the observational limit $\Phi = \Phi_{Fermi}$.  The cyan, blue, and magenta contours correspond to $\gamma_c = 1.3$,  $\gamma_c = 1.4$, and  $\gamma_c = 1.5$, respectively.}
    \label{fluxcomparedobserva}
\end{figure}

In Fig.~\ref{fluxcomparedobserva}, we show the observational limit as a function of the DM mass for the case of a depleted spike with spike radius given by Eq.~\ref{depletion}. The spike power law inside the spike radius is given by the adiabatic expectation, Eq.~\ref{gammaspeq}. The contours are the total integrated flux $\Phi$ in units of  $ \Phi_{Fermi} = 2.18 \times 10^{-8}$ photons/cm$^2$s in the energy range 1-100 GeV for a benchmark integrated photon count $N=1$. The annihilation cross section is parametrized by Eq.~\ref{eq:partialwave}, with $c_0 = 3 \times 10^{-26}$ cm$^{3}$ s$^{-1}$ and $c_1 = 1 \times 10^{-30}$ cm$^{3}$ s$^{-1}$, corresponding to a canonical thermal relic. The dashed line shows the observational limit $\Phi = \Phi_{Fermi}$. From bottom to top, the cyan, blue, and magenta contours correspond to $\gamma_c = 1.3$,  $\gamma_c = 1.4$, and  $\gamma_c = 1.5$, respectively.

We see that thermal relics with increasingly large masses are constrained for depleted spikes with increasing $\gamma_c$. 
In general, larger values of $\gamma_c$ lead to slightly steeper spike profiles inside the spike radius $r_{sp}$, however the dominant effect comes from $r>r_{sp}$ where $\gamma_c$ determines the flux.  Larger $\gamma_c$ leads to an increased integrated flux and higher mass reach.  For a given/determined $\gamma_c$, the sensitivity falls off as $1/m_\chi^2$ due to the decreasing number density of DM particles.

\subsection{Idealized Spike}

   \begin{figure}[ht]
  \centering
  {\includegraphics[width=\textwidth]{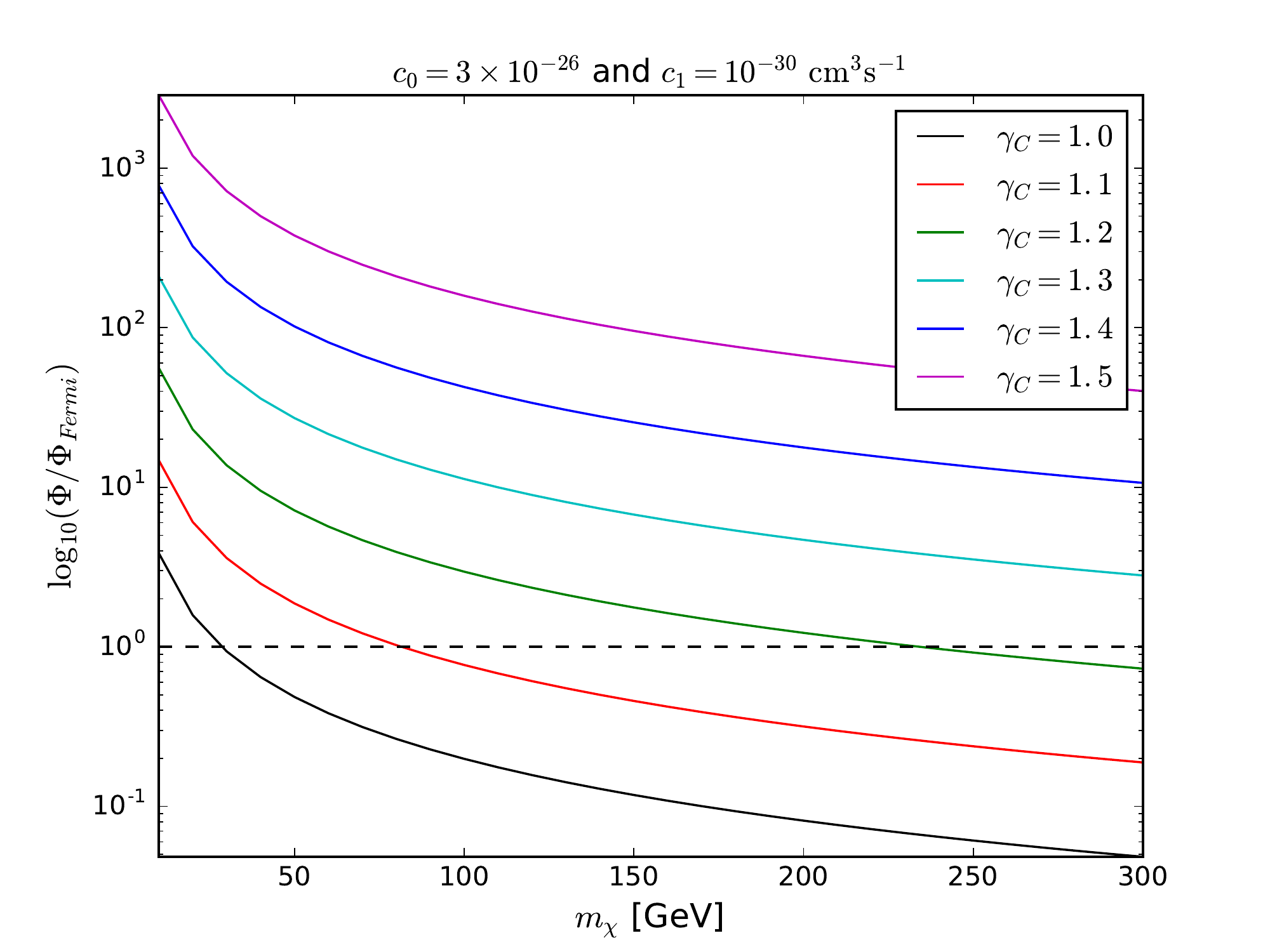}}
    \caption{\textbf{Idealized spike with} $\mathbf{\gamma_{sp} \, = \, \frac{9-2 \gamma_c}{4 - \gamma_c}}$  \textbf{: Observational reach versus mass plot - } The total integrated flux $\Phi$ in units of  $ \Phi_{Fermi} = 2.18 \times 10^{-8}$ photons/cm$^2$s coming from the source 3FGL J1745.6-2859c (Sgr A$^*$), in the energy range 1-100 GeV, and assuming an integrated photon count $N=1$. The annihilation cross section is parametrized by Eq.~\ref{eq:partialwave}, with $c_0 = 3 \times 10^{-26}$ cm$^{3}$ s$^{-1}$ and $c_1 = 1 \times 10^{-30}$ cm$^{3}$ s$^{-1}$, corresponding to a canonical thermal relic. The spike profile is given by Eq.~\ref{density}. The spike radius is given by the idealized case in Eq.~\ref{idealized}, with value $r_{sp} \sim 0.40$ pc. The spike power law inside the spike radius is given by Eq.~\ref{gammaspeq}, yielding values in the range $\gamma_{sp} \sim 2.3 \, - \, 2.4$. The dotted line shows the observational limit $\Phi = \Phi_{Fermi}$.  The  black, red, green, cyan, blue, and magenta contours correspond to $\gamma_c = 1.0$, $\gamma_c = 1.1$, $\gamma_c = 1.2$, $\gamma_c = 1.3$,  $\gamma_c = 1.4$, and  $\gamma_c = 1.5$, respectively.}
    \label{fluxcomparedobservb}
\end{figure}

Finally, we turn our attention to the possibility that the spike has not depleted as described by Eq.~\ref{depletion}, but rather remains intact or is described by a spike exponent different from the adiabatic expectation\footnote{A value of $\gamma_{sp}$ different from the adiabatic expectation of Eq.~\ref{gammaspeq} could arise at formation or through depletion over time, as briefly described in Sec.~\ref{sec:spikeprofile}.}.  We refer to both of these cases as {\it idealized}.

First, we address the case of a truly idealized spike; one that has not suffered depletion in any way and formed adiabatically.
In Fig.~\ref{fluxcomparedobservb}, we show the observational limit as a function of the DM mass for an idealized spike with spike radius given by Eq.~\ref{idealized} and its spike power law given by the adiabatic expectation, Eq.~\ref{gammaspeq}. The contours represent the total integrated flux $\Phi$ in units of  $ \Phi_{Fermi} = 2.18 \times 10^{-8}$ photons/cm$^2$s in the energy range 1-100 GeV for an integrated photon count $N=1$. The annihilation cross section is parametrized by Eq.~\ref{eq:partialwave}, with $c_0 = 3 \times 10^{-26}$ cm$^{3}$ s$^{-1}$ and $c_1 = 1 \times 10^{-30}$ cm$^{3}$ s$^{-1}$, corresponding to a canonical thermal relic. The dotted line shows the observational limit $\Phi = \Phi_{Fermi}$.  From bottom to top, the  black, red, green, cyan, blue, and magenta contours correspond to $\gamma_c = 1.0$, $\gamma_c = 1.1$, $\gamma_c = 1.2$, $\gamma_c = 1.3$,  $\gamma_c = 1.4$, and  $\gamma_c = 1.5$, respectively.

In Fig.~\ref{fluxcomparedobservb}, again, we see the $1/m_\chi^2$ dependence, but the most striking feature is clearly the much large mass reach in this case relative to the depleted case shown in Fig.~\ref{fluxcomparedobserva}.  Fig.~\ref{fluxcomparedobservb} shows that if the spike has suffered no depletion, then even very large DM masses are incompatible $\gamma_c \gtrsim 1.3$. Another way of reading this is that if the DM spike at our Galactic Center has not suffered much depletion, then the absence of a DM signal gives us an upper limit on $\gamma_c$.

  \begin{figure}[ht]
  \centering
  {\includegraphics[width=\textwidth]{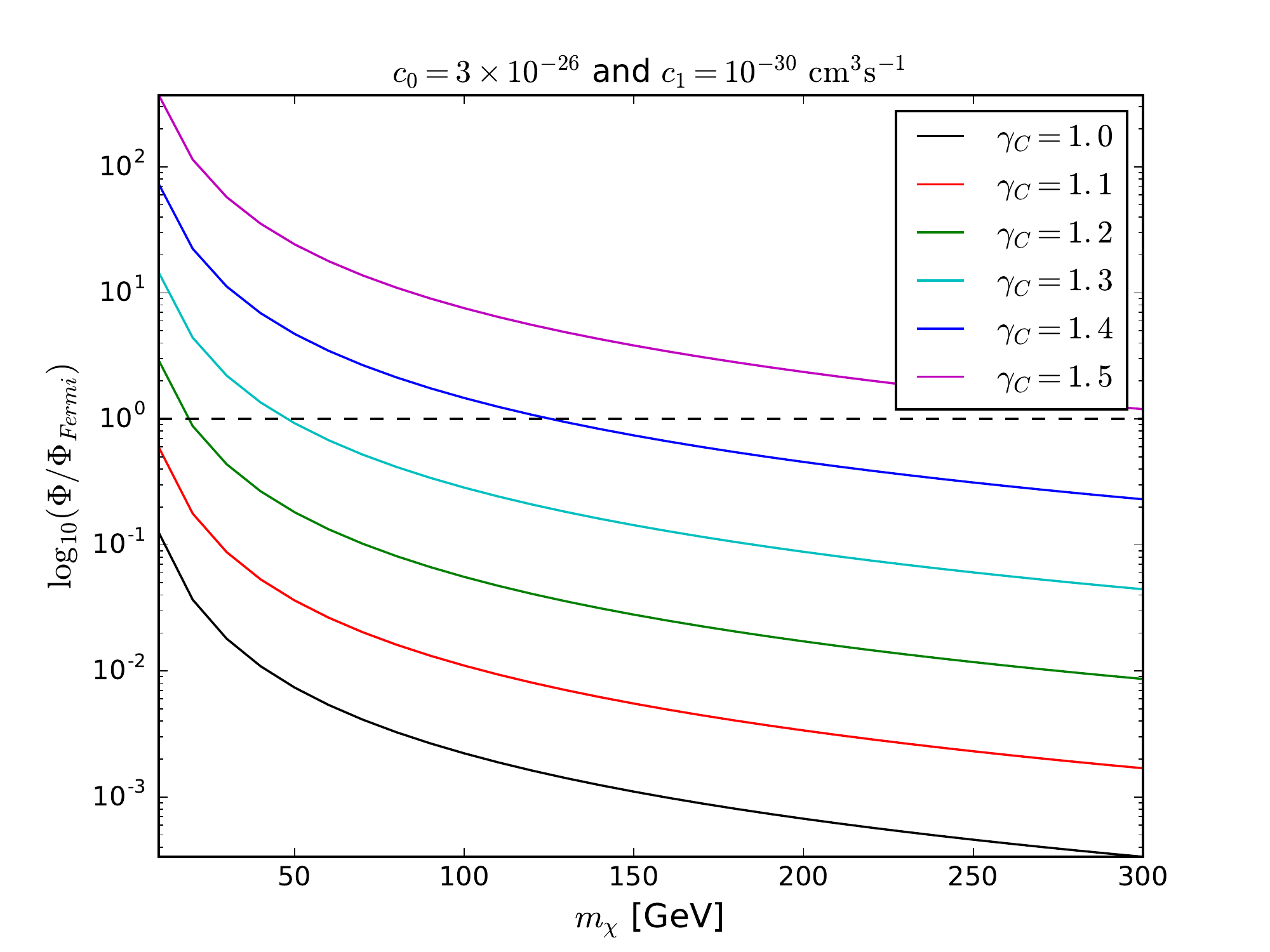}}
    \caption{\textbf{Idealized spike with} $\mathbf{\gamma_{sp} = 1.8}$ \textbf{: Observational reach versus mass plot - } The total integrated flux $\Phi$ in units of  $ \Phi_{Fermi} = 2.18 \times 10^{-8}$ photons/cm$^2$s coming from the source 3FGL J1745.6-2859c (Sgr A$^*$), in the energy range 1-100 GeV, and assuming an integrated photon count $N=1$. The annihilation cross section is parametrized by Eq.~\ref{eq:partialwave}, with $c_0 = 3 \times 10^{-26}$ cm$^{3}$ s$^{-1}$ and $c_1 = 1 \times 10^{-30}$ cm$^{3}$ s$^{-1}$, corresponding to a canonical thermal relic. The spike profile is given by Eq.~\ref{density}. The spike radius is given by the idealized case in Eq.~\ref{idealized}, with value $r_{sp} \sim 0.40$ pc. The spike power law inside the spike radius is given by $\gamma_{sp} = 1.8$. The dotted line shows the observational limit $\Phi = \Phi_{Fermi}$.  The black, red, green, cyan, blue, and magenta contours correspond to $\gamma_c = 1.0$, $\gamma_c = 1.1$, $\gamma_c = 1.2$, $\gamma_c = 1.3$,  $\gamma_c = 1.4$, and  $\gamma_c = 1.5$, respectively.}
    \label{fluxcomparedobservc}
\end{figure}

Lastly, in Fig.~\ref{fluxcomparedobservc} we show the observational limit as a function of the DM mass for the case of an idealized spike with spike radius given by Eq.~\ref{idealized}, but with $\gamma_{sp} = 1.8$.  One can view this as a different type of depletion, which may be from gravitational interactions with stars, or potentially from some other mechanism, however it is {\it idealized} in the sense that the spike radius, $r_{sp}$ given by Eq.~\ref{rbinit}, is unchanged over time. Other parameter choices are identical to Fig.~\ref{fluxcomparedobservb}. 
Obviously, since the spike exponent is smaller in this case than in the truly idealized spike shown in Fig.~\ref{fluxcomparedobservb}, the mass reach is substantially reduced. In fact, comparing Figs.~\ref{fluxcomparedobservc} and \ref{fluxcomparedobserva}, we see that flux from an idealized spike with $\gamma_{sp}=1.8$ is just a factor of a few larger than the flux from a depleted spike with the same $\gamma_c$.  In the future, these some-what degenerate cases may be resolved by carefully studying the extended spatial morphology (rather than just the point source flux) of a gamma-ray signal of DM annihilation.

Our results from this section may be summarized as follows: The degree to which the DM spike near the black hole can constrain DM models depends strongly on the parameters that determine the spike profile, such as the spike radius $r_{sp}$ and the parameters $\gamma_{sp}$ and $\gamma_c$ describing the  profile power-law behavior inside and outside the spike radius, respectively. Different choices of these parameters have been considered, ranked in order from the most conservative to the most optimistic:

$(i)$ A depleted spike with radius given by Eq.~\ref{depletion}, $\gamma_{c} = 1.0$, and $\gamma_{sp}$ given by Eq.~\ref{gammaspeq} ($\gamma_{sp}=\frac{9-2\gamma_c}{4-\gamma_c}$).  This is the most conservative choice of parameters we study, and the results are shown in the top left panels of Fig.~\ref{coc1resultsm100} and Fig.~\ref{coc1resultsm200}, for 100 GeV and 200 GeV DM, respectively. We see that for a 100 GeV DM candidate with annihilation cross section compatible with a thermal relic, this choice of spike parameters leads to a flux that is several orders of magnitude smaller than the current observational limit $\Phi_{Fermi}$.  Smaller values of $\gamma_{sp}$ would therefore also lead to unobservably small photon fluxes.

$(ii)$ A depleted spike with radius given by Eq.~\ref{depletion}, $\gamma_{c} = 1.1 \, - \,  1.5$, and $\gamma_{sp}$ given by Eq.~\ref{gammaspeq}. The mass reaches are plotted in Fig.~\ref{fluxcomparedobserva}, and the constraints in the DM annihilation plane $c_0$ and $c_1$ are plotted in the top right and bottom panels of Fig.~\ref{coc1resultsm100} and Fig.~\ref{coc1resultsm200}. It is clear that with increasing $\gamma_c$, the current observational limit $\Phi_{Fermi}$ can put some constraints on DM of various masses.

$(iii)$ An idealized spike with radius given by Eq.~\ref{rbinit}, $\gamma_{c} = 1.0 \, - \, 1.5$, and $\gamma_{sp} = 1.8$. The mass reaches are plotted in Fig.~\ref{fluxcomparedobservc}. Due to the larger spike radius, the reaches are generally greater than the depleted spike of Case $(ii)$, even with the smaller value of $\gamma_{sp}$.

$(iv)$ An idealized spike  with radius given by Eq.~\ref{rbinit}, $\gamma_{c} = 1.0 \, - \, 1.5$, and $\gamma_{sp} =  \, \frac{9-2 \gamma_c}{4 - \gamma_c}$. This is the most optimistic choice of parameters in the sense that it would imply a very prominent spike, and the results are shown in Fig.~\ref{fluxcomparedobservb}. 

In particular, it is clear that for a given value DM mass and $\gamma_{sp}$, a comparison between the depleted spike in Case $(i)$ and the idealized spike in Case $(iv)$ shows that the flux increases by a factor of $\sim \mathcal{O}(10^3)$ when one assumes that the spike radius remained at its idealized value. A comparison between Case $(iii)$ and Case $(iv)$ shows that for a given spike radius and DM mass, changing $\gamma_{sp}$ from 1.8 to the value predicted from Eq.~\ref{gammaspeq} (typically $2.3 \, - \, 2.4$) increases the flux by  $\sim \mathcal{O}(10 \, - \, 10^2)$. 

\section{Results: Constraints on Simplified Models} \label{simpresults}

In this section, we present a particular example that demonstrates the impact of the spike form on conclusions regarding the particle physics of DM interactions. Specifically, we describe the constraints that are obtained on simplified models of DM with $t$-channel mediators. For concreteness, we take the DM mass to be 100 GeV, and consider a subset of the parametrizations discussed in section \ref{results}.  As an example, we consider DM annihilating to $b \overline{b}$ final states.  We first describe this class of simplified models and give an overview of the calculation of the annihilation cross section, then we provide a discussion of the results.

\subsection{Simplified Model with $t$-Channel Mediators}

In this section, we describe some general features of models of DM that annihilate primarily through the $t-$channel. There is a vast amount of literature on these models, and we refer the reader to \cite{Garny:2015wea} and references therein.

For simplicity, we  focus on Majorana DM candidates $\chi$ with mass $m_{\chi}$ that couple to both left and right SM fermions $f_{L,R}$. The mediator sector consists of a pair of scalars $\widetilde{f}_{L,R}$, with a mixing angle $\alpha$ \cite{Sandick:2016zut}, \cite{Kumar:2016cum}. The standard case of mediator sectors coupling to right-handed SM fermions corresponds to the choice $\alpha = \pi/2$.

The interaction Lagrangian is given by
\begin{equation}
\label{eq:Lint}
  \mathcal{L}_{\text{int}}=\lambda_{L}\widetilde{f}_{L}^{\ast}\overline{\chi}P_{L}f+\lambda_{R}\widetilde{f}_{R}^{\ast}\overline{\chi}P_{R}f+\text{c.c.} \,\,,
\end{equation}
where the Yukawa couplings $\lambda_{L,R}$ may in general contain a $CP$-violating phase,
\begin{align}
  &\lambda_{L}\equiv\left|\lambda_{L}\right|e^{i\varphi/2}\,,& &\lambda_{R}\equiv\left|\lambda_{R}\right|e^{-i\varphi/2} \,\,.
\label{yukawasimplified}
\end{align}
Here, we set $\varphi=0$. The mixing angle $\alpha$ between the scalar mass and chiral eigenstates is given by
\begin{equation} \label{mixingangle}
  \left(\begin{array}{c}
    \widetilde{f}_{1} \\ \widetilde{f}_{2}
    \end{array}\right)
  =\left(
  \begin{array}{cc}
    \cos\alpha & -\sin\alpha \\
    \sin\alpha & \cos\alpha
  \end{array}\right)\left(
  \begin{array}{c}
    \widetilde{f}_{L} \\ \widetilde{f}_{R}
  \end{array}\right)\,.
\end{equation}
The two scalar mass eigenvalues are denoted as $m_{\widetilde{f}_{1}}$ and $m_{\widetilde{f}_{2}}$ in the following. There are thus the following free parameters in this class of simplified models \cite{Fukushima:2014yia}:
\begin{itemize}
\item the four masses, $m_{\chi}$, $m_{\widetilde{f}_{1}}$, $m_{\widetilde{f}_{2}}$ and $m_{f}$. 

\item the Yukawas $|\lambda_{L,R}|$, the scalar mixing angle $\alpha$, and the $CP$-violation phase $\varphi$ (here $\varphi=0$). 
\end{itemize}

In fact, this simplified model represents a slice of the parameter space of the Minimal Supersymmetric Standard Model (MSSM), in which bino DM couples to one generation of light sfermions. In the case of the MSSM, the Yukawa couplings are given by  
\bea
|\lambda_{L}|=\sqrt{2}g|Y_{L}| \nonumber \\
|\lambda_{R}|=\sqrt{2}g|Y_{R}| \,\, ,
\label{susyyukawa}
\eea
where $g$ is the electroweak coupling constant and the hypercharges are $|Y_{L}|=1/2$ and $|Y_{R}|=1$ for leptons and $|Y_{L}|=1/3$ and $|Y_{R}|=2/3$ for quarks. 

%

%
%

\begin{figure}[t]
  \centering
  {\includegraphics[width=.45\textwidth]{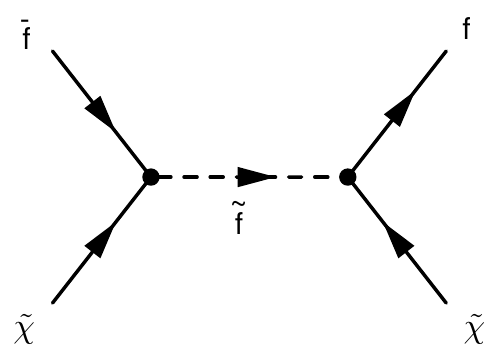}}\quad
  {\includegraphics[width=.45\textwidth]{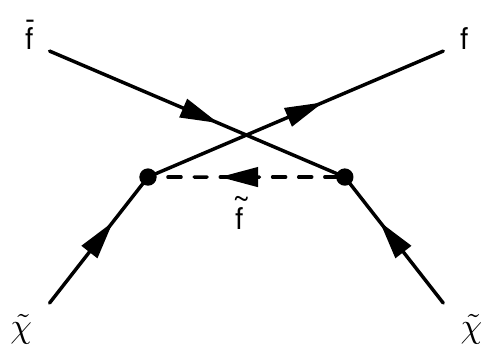}}
    \caption{Feynman diagrams for DM annihilation in the $t$-channel.}
    \label{feynanns}
\end{figure}

The relevant diagrams for DM annihilation in this model are given in Fig.~\ref{feynanns}.
Parametrizing the annihilation cross section in the standard way, the velocity-independent $s$-wave contribution $c_0$ is given in the limit $m_{f}/m_{\tilde {f}_i} \to 0$ by the simple expression
\begin{equation}
c_0 = {m_{\tilde \chi}^2 \over 2 \pi } g^4  Y_L^2 Y_R^2 \cos^2 \alpha \sin ^2 \alpha \left( {1 \over m_{\tilde {f}_1}^2
+ m_{\tilde \chi}^2} -{1 \over m_{\tilde {f}_2}^2  + m_{\tilde \chi}^2} \right) ^2 ,
\label{annc0}
\end{equation}

In the limit $m_{f}/m_{\tilde {f}_i} \to 0$,  the $v^2$-suppressed contribution, $c_1$ simplifies considerably, and the analytic expression is
\bea
 c_1= { m_{\tilde \chi}^2 \over 2 \pi } g^4 &\Biggl(& {(Y_L^4 \cos ^4 \alpha +Y_R^4 \sin ^4 \alpha )(m_{\tilde {f}_1}^4
 + m_{\tilde \chi}^4)\over (m_{\tilde {f}_1}^2 + m_{\tilde \chi}^2)^4} + { (Y_L^4 \sin ^4 \alpha +Y_R^4 \cos ^4 \alpha ) (m_{\tilde {f}_2}^4
 + m_{\tilde \chi}^4)\over (m_{\tilde {f}_2}^2  + m_{\tilde \chi}^2)^4}
\nonumber \\  &&
+ { 2(Y_L^4 +Y_R^4  )\sin ^2 \alpha  \cos ^2 \alpha (m_{\tilde {f}_1}^2 m_{\tilde {f}_2}^2 + m_{\tilde \chi}^4)\over (m_{\tilde {f}_1}^2
+ m_{\tilde \chi}^2)^2 (m_{\tilde {f}_2}^2  + m_{\tilde \chi}^2)^2}
\nonumber \\  &&
+ { Y_L^2 Y_R^2  \sin ^2 \alpha  \cos ^2 \alpha(m_{\tilde {f}_1}^2 - m_{\tilde {f}_2}^2 )^2\over 2(m_{\tilde {f}_1}^2
+ m_{\tilde \chi}^2)^4 (m_{\tilde {f}_2}^2  + m_{\tilde \chi}^2)^4}
 \Bigl[ 3 m_{\tilde {f}_1}^4 m_{\tilde {f}_2}^4 - 52 m_{\tilde \chi}^4 m_{\tilde {f}_1}^2 m_{\tilde {f}_2}^2 + 3 m_{\tilde \chi}^8
\nonumber \\  &&
- 14m_{\tilde \chi}^2(m_{\tilde {f}_1}^2 + m_{\tilde {f}_2}^2)( m_{\tilde \chi}^4+ m_{\tilde {f}_1}^2 m_{\tilde {f}_2}^2)
-5m_{\tilde \chi}^4(m_{\tilde {f}_1}^4 + m_{\tilde {f}_2}^4)  \Bigr]
\Biggr) .
\label{annc1}
\eea
 We note that the velocity-suppressed terms arise from both $s$-wave and $p$-wave matrix elements. We also note that $c_0$ and $c_1$ do depend on $\varphi$ in terms proportional to $m_f$.  Additionally, these $m_f$-dependent terms carry coefficients involving $Y_L$ and $Y_R$ such that there can be interesting cancellations, even in $c_0$. In our results, we will use the full expressions for $c_0$ and $c_1$, including $m_f$-dependent terms.  

%
%



\subsection{Constraints on Simplified Models with $t$-Channel Mediators}

We now discuss the constraints in the context of the simplified model introduced above. We adapt Fig.~\ref{coc1resultsm100}, which gives contours of the flux $\Phi$ in units of the current observational limit $\Phi_{Fermi}$ in the $(c_0, c_1)$ plane, and overlay a scan over the parameters of our simplified model. We consider two representative cases: the case of a depleted spike with $\gamma_c = 1.3$, and the case of an idealized spike with $\gamma_c = 1.0$. For each case, we scan over the mixing angle $\alpha$. 

  \begin{figure}[ht]
  \centering
  {\includegraphics[width=.48\textwidth]{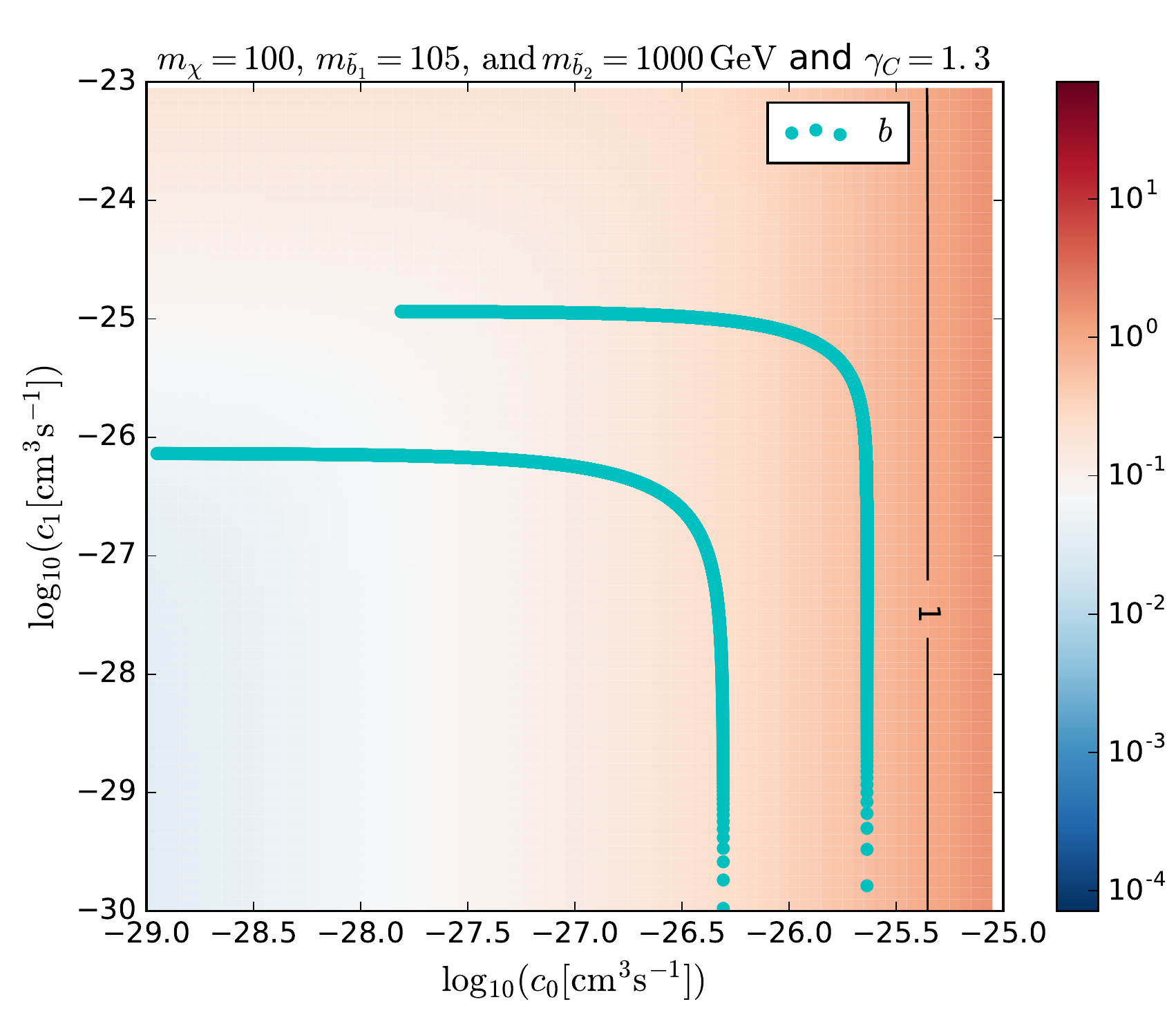}}\hspace{2mm}
  {\includegraphics[width=.48\textwidth]{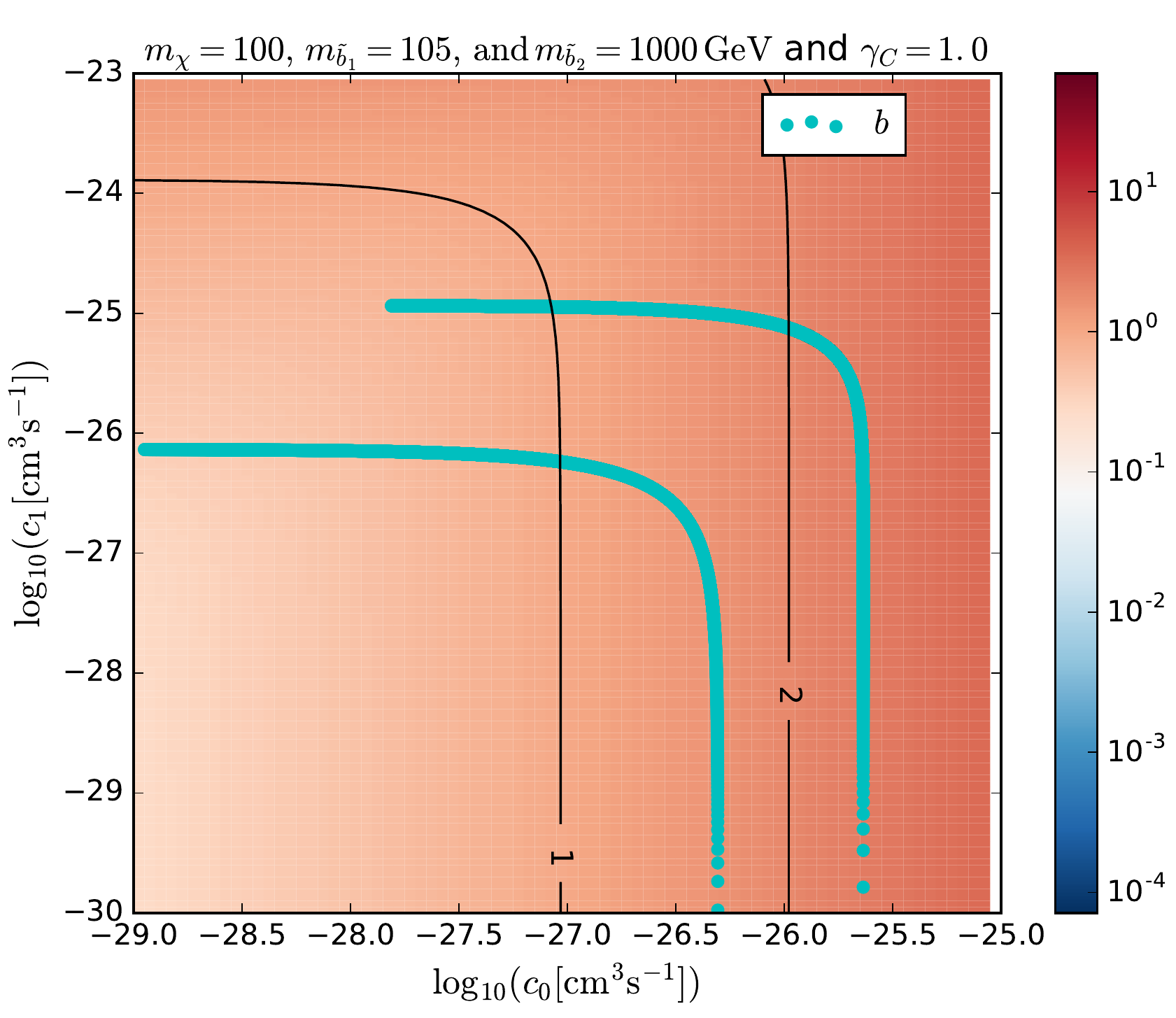}}
    \caption{\textbf{Depleted/Idealized spike, Simplified model - scan over mixing angle} $\mathbf{\alpha}$:  In the left panel we present the constraints under the assumption of a depleted spike, and in the right panel we assume an idealized spike.  In each case, the green points correspond to a scan over the mediator mixing angle $\alpha$ defined by Eq.~\ref{mixingangle}. The Yukawa couplings are held fixed at their supersymmetric values, given by Eq.~\ref{susyyukawa}. The DM mass is 100 GeV and the lightest sbottom mass is 105 GeV, with all other superpartners heavy.  The solid black line denotes the contour of the integrated flux $\Phi_{Fermi} = 2.18 \times 10^{-8}$ photons/cm$^2$s, coming from the source 3FGL J1745.6-2859c (Sgr A$^*$), assuming an energy range of 1-100 GeV and $b \overline{b}$ final states in DM annihilation.   The spike profile is given by Eq.~\ref{density}. In the left panel, the spike radius is given by the \textbf{depleted} case in Eq.~\ref{depletion} and $\gamma_c  = 1.3$,  yielding $\gamma_{sp} \sim 2.37$.  In the right panel, the spike radius is given by the \textbf{idealized} case in Eq.~\ref{idealized} and $\gamma_c  = 1.0$, yielding $\gamma_{sp} \sim 2.33$.
    \label{simpc0c1}}
\end{figure}


\begin{figure}[ht]
  \centering
  {\includegraphics[width=\textwidth]{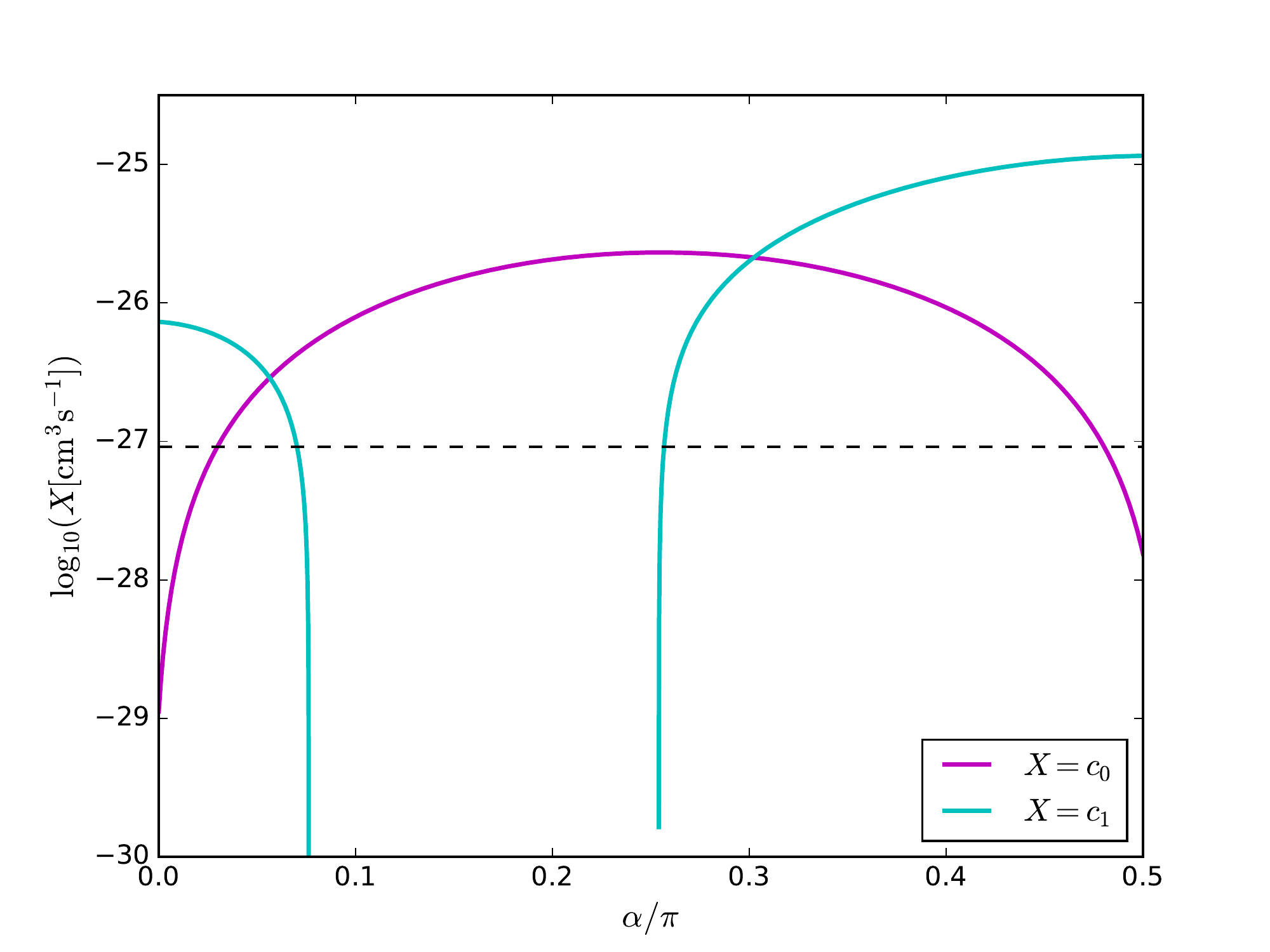}}
    \caption{ \textbf{Dependence of} $\mathbf{c_0}$ and $\mathbf{c_1}$ \textbf{on} $\mathbf{\alpha}$: The purple and blue curves show the dependence of $c_0$ and $c_1$ on $\alpha$ as obtained from Eq.~\ref{annc0} and Eq.~\ref{annc1}, respectively. The horizontal dotted line corresponds to the contour $\Phi = \Phi_{Fermi}$ from Fig.~\ref{simpc0c1aidealalp} (idealized spike with $\gamma_c  = 1.0$). Values of $c_0$ above the dotted line are constrained by the integrated flux of photons coming from the source 3FGL J1745.6-2859c (Sgr A$^*$).}
    \label{alpdepc0c1}
\end{figure}

In the left and right panels of Fig.~\ref{simpc0c1}, we consider the case of a depleted spike and idealized spike, respectively. The cyan dots show a scan over the mixing angle $\alpha$  defined by Eq.~\ref{mixingangle}, holding the Yukawa couplings fixed at their supersymmetric values given in Eq.~\ref{susyyukawa}. The scan is performed in the range $\alpha \, = \, 0$ to $ \pi/2$, which traces out a boomerang in the plane.  The low-energy spectrum of the SUSY model we consider consists of bino DM with mass $m_{\chi} = 100$ GeV, the lightest bottom squark with mass $m_{\tilde{b}} = 105$ GeV, and all other superpartners heavy. The solid black line denotes the contour of the integrated flux $\Phi_{Fermi} = 2.18 \times 10^{-8}$ photons/cm$^2$s. 

From the depleted case with $\gamma_c = 1.3$ considered in the left panel of Fig.~\ref{simpc0c1}, we can see that current observational constraints just barely begin to constrain the parameter space. For $\gamma_c = 1.0$, as one would expect, the results are even weaker, while we have checked that the case of $\gamma_c = 1.5$ constrains a significant portion of the parameter space. Indeed, the constraints are much stronger for the case of an idealized spike, shown in the right panel of  Fig.~\ref{simpc0c1}. We can see that even for $\gamma_c = 1.0$ in the case of an idealized spike, the current observational limits constrain a large part of the parameter space.
 
The resulting constraints on $\alpha$ are displayed in Fig.~\ref{alpdepc0c1}. The magenta and cyan curves show the dependence of $c_0$ and $c_1$ on $\alpha$ as obtained from Eq.~\ref{annc0} and Eq.~\ref{annc1}, respectively. For $\alpha \, \approx \, 0 , \, \pi/2$, the annihilation cross section drops precipitously since the contribution from $c_0$ suffers from chiral suppression and the contribution from $c_1$ is velocity-suppressed. These are the regions where the scans in Fig.~\ref{simpc0c1} are cut off towards the left, where $c_0$ becomes small. Conversely, there is a range of values $\alpha \, \approx \, 0.08 \pi \, - \, 0.25 \pi$, where $c_1$ becomes small, but $c_0$ remains large. These are the regions that are cut off towards the bottom of the scans in Fig.~\ref{simpc0c1}, where $c_1$ is small. 

The horizontal dotted line in Fig.~\ref{alpdepc0c1} corresponds to $c_0 \, \approx \,  10^{-27}$ cm$^3$ s$^{-1}$, which is where the $\Phi = \Phi_{Fermi}$ contour for the idealized case in the right panel of Fig.~\ref{simpc0c1} intersects the $c_0$ axis. Values of $c_0$ larger than this yield an integrated photon flux that is constrained by the source 3FGL J1745.6-2859c (Sgr A$^*$). Thus, from Fig.~\ref{alpdepc0c1}, it is clear that either $\alpha \approx 0$ or $\alpha \approx \pi/2$ if the spike is idealized. Very different conclusions are reached if the spike is depleted.

While this simplified model describes a subset of the MSSM parameter space, it need not be confined to the MSSM.  For example, it is possible that the Yukawa couplings, $\lambda_{L,R}$, deviate from their MSSM values.  In the absence of a signal, one could then constrain the couplings $\lambda_{L,R}$ for any combination of new particle masses and mixings.  If the form of the spike is understood, using the point source flux to constrain the model parameters could be a powerful technique.  Alternatively, as will be explored in the next section, if we have some indication of the DM model, then the point source flux could help us understand the spike morphology, and therefore provide a window into the astrophysics of the very central region of our Galaxy.



%
%

\section{Constraints on Spike Parameters from DM Annihilations} \label{spikeconstras}

In this Section, we invert the approach we have hitherto taken to demonstrate the potential power of gamma-ray observations of a known DM candidate to determine the spike profile (and potentially learn something about the astrophysics that led to it). Although the most recent analysis indicates that the excess of GeV photons from the Galactic Center region observed by Fermi-LAT is most likely not due to DM~\cite{TheFermi-LAT:2017vmf}, it is instructive to take this case as an example. We calculate the constraints on the spike parameters in our model under the assumption of a particular DM model designed to explain the excess of $\sim 1-3$ GeV gamma-rays from the Galactic Center. Specifically, we take as our benchmark point 
\bea
m_{\chi} \, = \, 49 \,\, {\rm GeV} \nonumber \\ 
c_0 \, = \, 1.76 \times 10^{-26} \, {\rm cm}^{3} {\rm s}^{-1} \nonumber \\
c_1 \, = \, 1.0 \times 10^{-30} \, {\rm cm}^{3} {\rm s}^{-1}, \,\,
\label{GCcandidate}
\eea
and assume $b \overline{b}$ final states, as in \cite{TheFermi-LAT:2015kwa}, \cite{Karwin:2016tsw}, \cite{Daylan:2014rsa}.

Clearly, many choices for the spike parameters and the relationships among them exist, and considering different combinations would lead to different kinds of constraints on the parameter space. As a representative case, we consider a depleted spike and put constraints on the $\gamma_{sp}$ vs. $\gamma_c$ plane. The spike radius is given by Eq.~\ref{depletion}, and though we do not explicitly enforce the adiabatic relation for $\gamma_{sp}$, we do plot it as a dashed line in the plane.

   \begin{figure}[ht]
  \centering
  {\includegraphics[width=\textwidth]{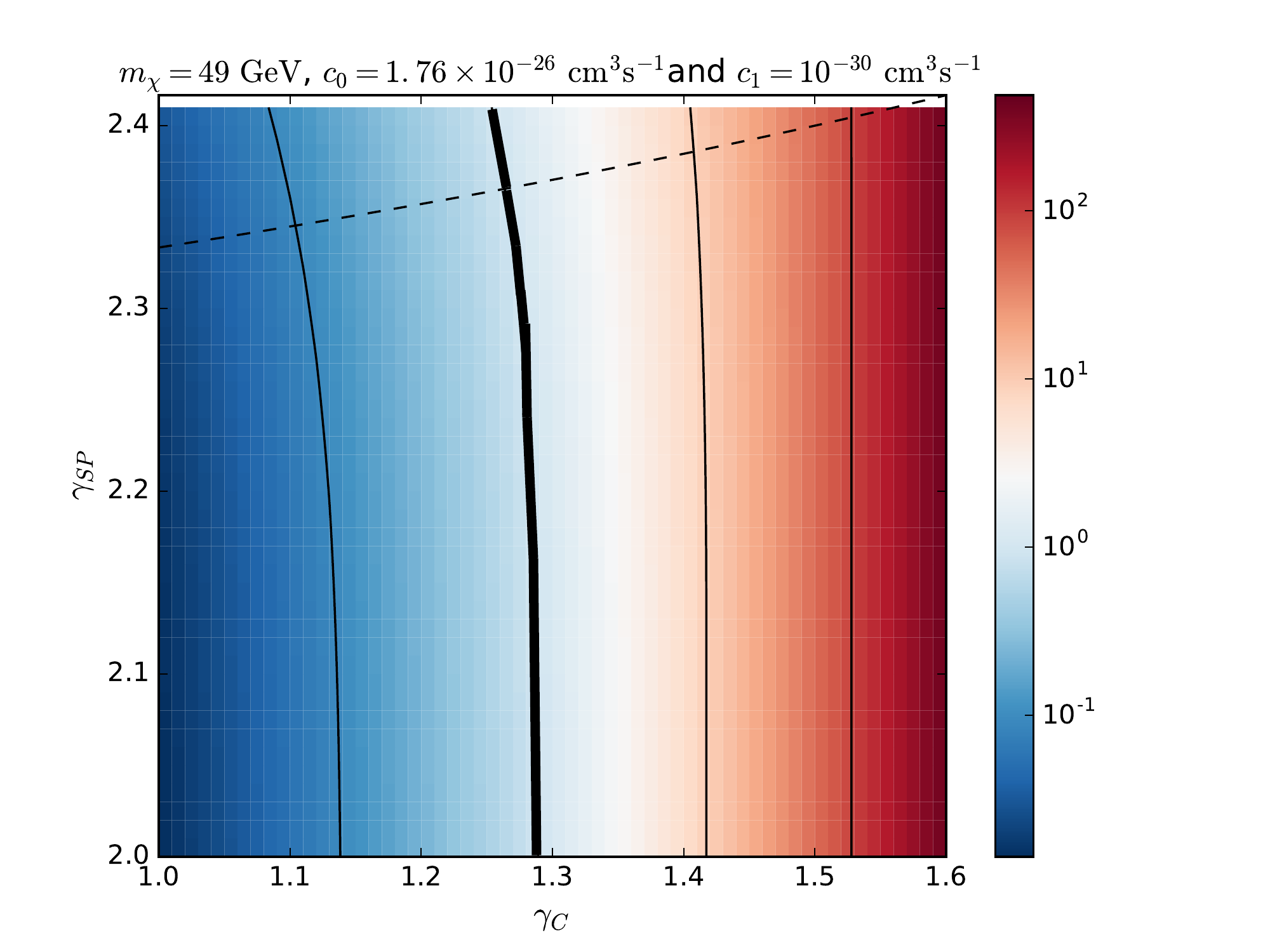}}
    \caption{\textbf{Constraints on Spike Parameters, assuming Depleted Spike and GC Excess:} The DM mass is 49 GeV and the annihilation cross section is parametrized by Eq.~\ref{eq:partialwave}, with $c_0 \, = \, 1.76 \times 10^{-26} \, {\rm cm}^{3} {\rm s}^{-1}$ and $c_1 \, = \, 1.0 \times 10^{-30} \, {\rm cm}^{3} {\rm s}^{-1}$. The solid black contours denote the integrated flux $\mathbf{\Phi}$ in units of $ \Phi_{Fermi} = 2.18 \times 10^{-8}$ photons/cm$^2$s coming from the source 3FGL J1745.6-2859c (Sgr A$^*$), assuming an energy range of 1-100 GeV and $b \overline{b}$ final states in DM annihilation. The bold contour corresponds to $\Phi = \Phi_{Fermi}$  The spike profile is given by Eq.~\ref{density}. The spike radius is given by the depleted case in Eq.~\ref{depletion}, with values $r_{sp} \sim 0.002 \, - \, 0.046$ pc. The dotted line shows the relation between $\gamma_{sp}$ and $\gamma_c$ given by Eq.~\ref{gammaspeq}.}
    \label{GCflux}
\end{figure}

The results are displayed in Fig.~\ref{GCflux}. The solid black contours denote the integrated flux $\mathbf{\Phi}$ in units of $ \Phi_{Fermi} = 2.18 \times 10^{-8}$ photons/cm$^2$s coming from the source 3FGL J1745.6-2859c (Sgr A$^*$), assuming an energy range of 1-100 GeV. The dashed line shows the adiabatic relation between $\gamma_{sp}$ and $\gamma_c$ given by Eq.~\ref{gammaspeq}. It is clear that for a depleted spike, $\gamma_c \, \gtrsim \, 1.3$ is incompatible with a DM interpretation of the Galactic Center excess for most values of $\gamma_{sp}$.  This is even true for very steep spikes with large $\gamma_{sp}$; as long as $\gamma_c$ is not too large, these scenarios are not excluded by the point source flux.

Additionally, the fact that the contours are nearly independent of $\gamma_{sp}$, i.e.~mostly vertical, indicates that it is not actually the spike that is responsible for the bulk of the photons.  Instead, the spike is actually fairly insignificant relative to the smooth component of the halo.  Ultimately, with some knowledge of the properties of DM, perhaps an observed, or unobserved, flux may help us learn about the DM profile near the Galactic Center, and possibly even the astrophysical mechanisms at play.


   \begin{figure}[ht]
  \centering
  {\includegraphics[width=\textwidth]{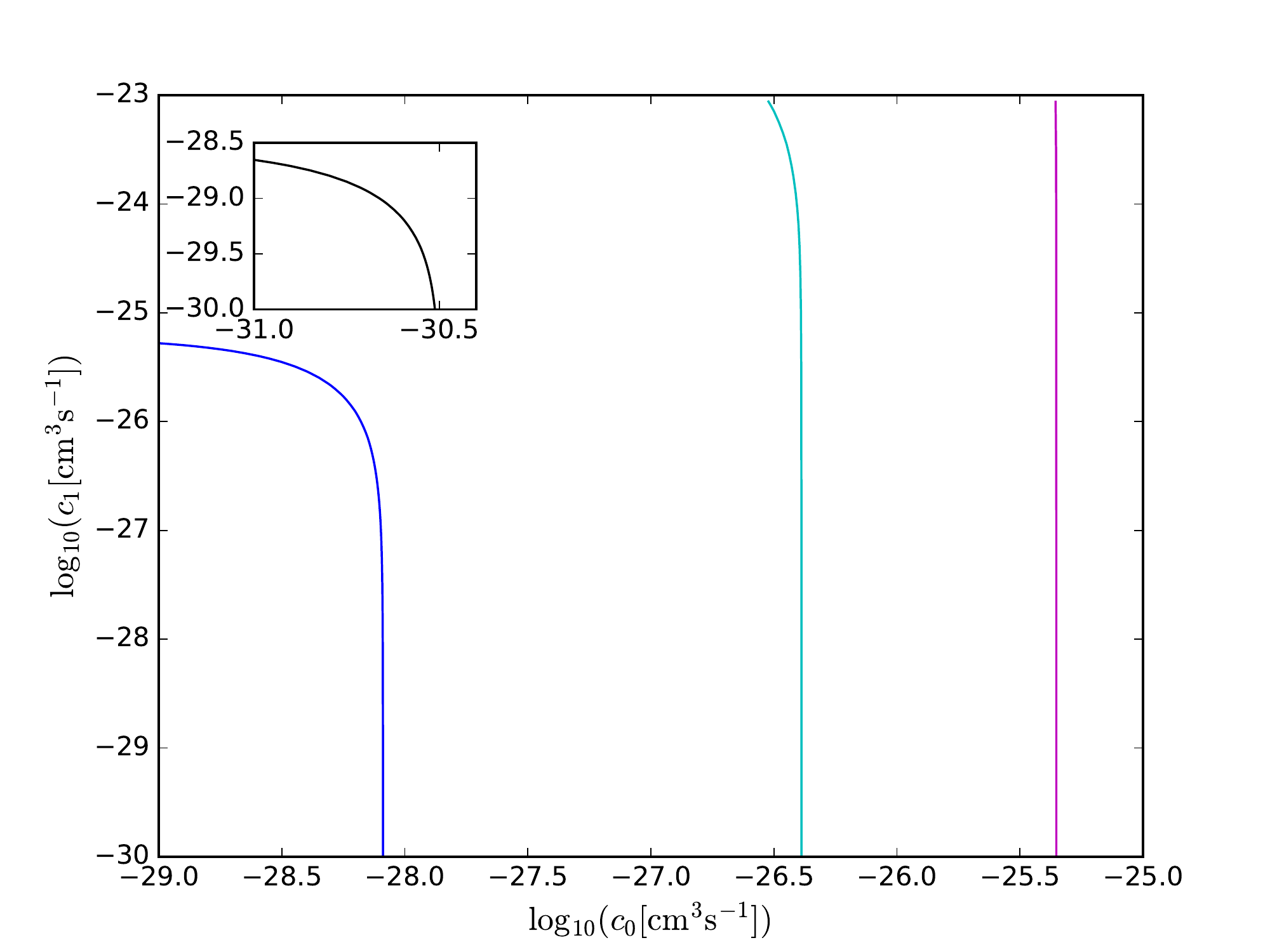}}
    \caption{\textbf{Constraints on DM:} The DM mass is 49 GeV and the annihilation cross section is parametrized by Eq.~\ref{eq:partialwave}. The contours denote the cases where the integrated flux $\Phi =  \Phi_{Fermi} = 2.18 \times 10^{-8}$ photons/cm$^2$s coming from the source 3FGL J1745.6-2859c (Sgr A$^*$), assuming an energy range of 1-100 GeV and $b \overline{b}$ final states in DM annihilation.   The spike profile is given by Eq.~\ref{density}. The spike radius is given by the depleted case in Eq.~\ref{depletion}, with values $r_{sp} \sim 0.002 \, - \, 0.046$ pc. The magenta, cyan, and blue contours correspond to $\gamma_c = 1.3, \, 1.4, \, 1.5$, respectively. The inset shows the contour corresponding to $\gamma_c = 1.6$. }
    \label{GCfluxc0c1plane}
\end{figure}

In Fig.~\ref{GCfluxc0c1plane}, we display the constraints on a DM candidate with a mass of 49 GeV, but allowing the coefficients $c_0$ and $c_1$ as free parameters. The contours denote the cases where the integrated flux $\Phi =  \Phi_{Fermi} = 2.18 \times 10^{-8}$ photons/cm$^2$s.  The magenta, cyan, and blue contours correspond to $\gamma_c = 1.3, \, 1.4, \, 1.5$, respectively. The inset shows the contour corresponding to $\gamma_c = 1.6$.  If $\gamma_c$ is very large, then the DM annihilation cross section must be very small indeed, to avoid overproducing the GC point source gamma-ray flux.

\section{Conclusions}

In this paper, we have studied contributions of a DM spike near the central black hole of our Galaxy to the gamma-ray flux $\Phi$. As our reference gamma-ray source, we have taken 3FGL J1745.6-2859c (Sgr A$^*$) from Fermi-$LAT$'s Third Point Source Catalog. We have taken into account a variety of astrophysical parameters describing the spike, and calculated the resulting constraints on general models of DM. We have then taken these constraints and applied them to a specific simplified model of fermionic DM with $t-$channel mediators. Finally, we have inverted our approach and considered the case of a DM candidate fitting the Galactic Center excess, and calculated the resulting constraints on the space of astrophysical spike parameters. 

We have found that the spike formation history and profile parameters have a profound effect on the extent to which models of DM can be constrained. 

$(i)$ For the most conservative choice of parameters (a depleted spike with radius given by Eq.~{depletion}, $\gamma_{c} = 1.0$), the flux for a 100 GeV thermal relic is several order of magnitude below current observational limits. We have then considered a series of less conservative choices. 

$(ii)$ A depleted spike with steeper cusp profile can constrain thermal relics of different masses depending on $\gamma_c$, as shown in Fig.~\ref{fluxcomparedobserva}. We see that thermal relics, approximately of masses 15 GeV, 50 GeV, and 140 GeV, are constrained by the choice of spike profile and different selections of $\gamma_c = 1.3, \, 1.4, \,$ and 1.5, respectively.

$(iii)$ An idealized spike which has not undergone attenuation improves the results considerably; the mass reach is shown in Fig.~\ref{fluxcomparedobservb}. This assumes that the inner spike profile corresponds to a scenario where the DM spike formed in response to the adiabatic growth of the black hole, i.e., $\gamma_{sp} \sim 2.3 - 2.4$. We see that thermal relics, approximately of masses 25 GeV, 80 GeV, and 240 GeV, are constrained by the choice of spike profile and different selections of $\gamma_c = 1.0, \,\, 1.1,$ and 1.2, respectively.

$(iv)$ Relaxing the assumption of an adiabatic growth of the black hole results in less steep spike profiles; for a particular choice of smoother profile $\gamma_{sp} = 1.8$, the mass reach is shown in Fig.~\ref{fluxcomparedobservc}. We see that thermal relics, approximately of masses 15 GeV, 50 GeV, and 140 GeV, are constrained by the choice of spike profile and different selections of $\gamma_c = 1.2, \,\, 1.3,$ and 1.4, respectively.

We have then gone on to apply these results for the simplified model of fermionic DM with $t$-channel mediators described by Eq.~\ref{eq:Lint}. In particular, we have performed scans over the mixing angle $\alpha$ and the Yukawa couplings of the theory, and checked to what extent the models are constrained by the observational limits of the gamma-ray flux from 3FGL J1745.6-2859c (Sgr A$^*$). We have found that while a depleted spike radius just barely begins to constrain the parameter space, an idealized spike constrains large parts of it, even for the most conservative choice of the cusp profile $\gamma_c = 1.0$.

Furthermore, we explored the possibility of constraining the space of astrophysical spike parameters, assuming that we know something about the properties of the DM, taking as an example a proposed DM candidate to explain the excess of GeV photons from the Galactic Center observed by Fermi-LAT.  If the spike is depleted, we find that moderate values of $\gamma_c \, \lesssim \, 1.3$ would be compatible with this particular model of DM for most values of $\gamma_{sp}$, but some values of $\gamma_{c}$ could certainly be excluded. 

Finally, we'd like to note that the depletion we assume is for a heating timescale of $10^9$ yr, which may be either shorter or longer than is realized in nature.  If depletion is less strong, which here might be realized by a longer heating timescale, then the fluxes from any given model would be larger.  This means that the power to exclude DM models would be greater, or, conversely, the power to use some knowledge about the properties of DM to constrain $\gamma_{sp}$ and $\gamma_c$ would be greater than in the depleted scenarios presented here.

\section{Acknowledgement}

We would like to thank Mustafa Amin for collaboration in the early stages of this work. PS is supported in part by NSF Grant No. PHY-1417367.

\end{document}